\documentclass[11pt]{article}
\usepackage{bbm}
\usepackage{amsfonts}
\usepackage{amssymb,amsmath,amsthm,amsfonts}
\usepackage{graphicx}
\usepackage[normal]{subfigure}
\theoremstyle{plain}

\newtheorem{proposition}{Proposition}[section]

\theoremstyle{definition}
\begin{document}
\title{{\bf Impact of heterogenous prior beliefs and disclosed
insider trades
\thanks{For useful discussions, we thank Deqing Zhou and Bowen Shi.
We also would like to thank Jiaan Yan, Jianming Xia and other
seminar participants at the Institute of Applied Mathematics,
 Academy of Mathematics and Systems Science,
Chinese Academy of Sciences. The first named author is grateful for
financial support for National Natural Science Foundation of China
(No.10721101) and China's National 973 Project (No.2006CB805900).
The second author  also would like to thank Hengfu Zou, Zhixiang
Zhang and other seminar participants at Central University of
Finance and Economics. }}}
\author{Fuzhou Gong$^1$ \quad
  Hong Liu$^{2,3}$ \thanks{Corresponding author. E-mail: mathhongliu@gmail.com, liuh653@nenu.edu.cn.}
  \\ {\small  $^{1}$ Institute of Applied Mathematics, Academy of
Mathematics and Systems Science, }
\\ {\small Chinese Academy of Sciences,
Beijing, 100190, P.R. China.}
\\{\small  $^{2}$MOE Key Laboratory of
Applied Statistics, School of Mathematics and Statistics,}
%Ministry of Education,}
\\{\small  Northeast Normal University, Changchun, 130024, P.R. China}
\\ {\small  $^{3}$China Economics and Management Academy, and CIAS,}
\\ {\small Central University of Finance and Economics, Beijing, 100081, P.R. China}
\date{}}
\maketitle

 \vskip .3cm \noindent

\begin{abstract}

In this paper, we present a multi-period trading model by assuming
that traders face not only asymmetric information but also
heterogenous prior beliefs, under the requirement that the insider
publicly disclose his stock trades after the fact. We show that
there is an equilibrium in which the irrational insider camouflages
his trades with a noise component so that his private information is
revealed slowly and linearly whenever he is overconfident or
underconfident. We also investigate the relationship between the
heterogeneous beliefs and the trade intensity in the presence of
trade disclosure, and show that  the weights on asymmetric
information and heterogeneous prior beliefs are opposite in sign and
they change alternatively in the next period. Under the requirement
of disclosure, the irrational insider trades more aggressively and
leads to smaller market depth. Moreover, the co-existence of
``public disclosure requirement'' and ``heterogeneous prior
beliefs'' leads to the fluctuant multi-period expected profits and a
larger total expected trading volume which is positively related to
the degree of heterogeneity. More importantly, even public
disclosure may lead to negative   profits of the irrational
insider's in some periods, inside trading remains profitable from
the whole trading period.

 \vspace{0.3cm}

\vspace{0.3cm}


\end{abstract}

\section{Introduction}
\setcounter{equation}{0}

\quad\quad Anybody, who has a casual look at the security  or stock
markets, would be amazed by the fluctuation features of financial
markets, such as the fluctuation of the trading volume, the price
and the short-term aggressive trading. But these fascinating
features have rarely been studied in  theoretical finance, perhaps
for the reason that they seem close to the irrational behavior of
traders and can not be explained in the rational expectations. In
this paper, we try to study these fascinating features of the market
and address the implications on financial market of irrational
behavior, by developing a trading model to characterize the optimal
behavior of an irrational insider who possesses long-lived private
information about the fundamental value of a security under
mandatory disclosure requirements.

In his pioneering and insightful paper,
 Kyle (1985) introduces a dynamic model of insider trading where an insider
receives only one signal and the fundamental asset value does not
change over time. Through trade, the insider progressively releases
his private information to the market as he exploits his
informational advantage. Kyle (1985) also points out that liquidity
trading provides camouflage which conceals informed trading such
that  the informed trading is swamped by liquidity trading. Based on
Kyle (1985), Huddart, Hughes and Levine (2001) consider the case of
the same insider under disclosure requirement as mandated by US
securities laws. The ex-post disclosure of the insider's trades
changes the equilibirum strategy of the insider, given that market
makers can infer information from the insider's previous trade
before the next round of trading. Huddart, Hughes and Levine (2001)
find that public disclosure of the insider's traders nevertheless
accelerates the price discovery process and lowers trading costs by
comparison to the case with no disclosure requirement. Zhang (2004)
extends Huddart, Hughes and Levine (2001) by incorporating the
condition that the monopoly insider is risk-averse, and he finds
that under disclosure requirements, the risk-averse insider is more
concerned about the risk of sub-optimally revealing his information
by mandatory disclosure and his private information is revealed
slowly. Also, Zhang (2008) analyzes a dynamic market where outsiders
share part of the information about a security with a corporate
insider and update their incomplete information by learning from
disclosed insider
 trades.  Recently, Gong and Liu (2011) study the optimal  behaviors of
 competitive insiders and their influences to the market under disclosure requirements.
 All the papers mentioned above have the assumption that the
 insider is rational. An interesting question is what the behavior
 of  an irrational insider behavior under mandatory disclosure
 requirements.

 In the past few years, lots of literatures have studied the behaviors
 of irrational traders and their influences to the market. Odean (1998) presents three
 different markets structures, two of which examine price-taking overconfident informed
 traders, and one which looks as  Kyle (1985)'s  setting. His results depend a
 lot on the specific assumptions about
 risk preferences of agents and competition.  Benos (1998)
  explicitly models investor behavior in financial markets allowing
  for traits linked to a notion of imperfect rationality. He studies
  an extreme form of posterior over confidence where some risk
  neutral investors overestimate the precision of their private
  information. In addition, Kyle and Wang (1997) and Wang  (1997)
consider the case of heterogeneous prior
 beliefs. The former considers heterogeneous prior beliefs in their
study of the survival of irrational traders in a duopoly context
while the later examines the implication of overconfidence for
delegated fund management in both learning and evolutionary game
models.    Harris and Raviv (1993) and  Wang  (1998) use
heterogeneous prior beliefs to explain the enormity of volume traded
each day. Specifically, Wang  (1998) extends the model of Kyle
(1985) by incorporating heterogeneous prior beliefs. He finds that
in equilibrium, the informed trader, facing both asymmetric
information and heterogeneous prior beliefs, smoothes out his
trading on asymmetric information gradually over time, but
concentrates his entire trading on heterogeneous beliefs toward the
last few periods. Since under US securities laws insiders associated
with a firm must report to the Securities and Exchange Commission
trades they make in the stock of that firm, we try to study the
optimal behaviors of an irrational insider, and address its impact
to the market under the mandatory disclosure requirements in this
paper, using the framework of Kyle (1985).

Kyle (1985)'s model has been widely used to analyze financial market
microstructure and the value of information. For example, Holden and
Subrahmanyan (1992) and Foster and Viswanathan (1996) consider a
market with multiple the competing insiders, and they show  that
competition among insiders accelerates the release of their private
information. Back (1992) formalizes  and extends  the model by
showing the existence of a unique equilibrium beyond the
Gaussian-linear framework. Remarkable, when the asset value has a
log-normal distribution, the price process becomes a geometric
Brownian motion as is usually assumed in finance. Holden and
Subrahmanyan (1994) assume  the single informed trader is
risk-averse. They show that both monopolistic and competing informed
traders choose to exploit rents rapidly, causing market depth to be
low in the initial periods and high in later periods and causing
information to be revealed rapidly, unlike in the case  of a
risk-neutral monopolist considered by Kyle (1985). Gong and Zhou
(2010) improve the Kyle (1985) model by loosing the assumption of
constant pricing rule and give a new framework to analysis the
insider's behavior.\footnote{We will consider this problem using the
new framework in other papers, and the purpose of this paper is to
analysis the impact of heterogenous prior beliefs and disclosed
insider trades to the market using the framework of Kyle (1985) so
as to
 compare with the modified model}. Also,  Fishman and Hagerty (1992), Luo (2001), Rochet and
Vila (1994), and Jain and Mirman (1999) e.t. have used variance of
Kyle's model to analyze and to explain real financial phenomena.

We consider a model  in which traders face both asymmetric
information and heterogeneous prior beliefs, with the requirement
that  the insider publicly disclose his stock trades after the fact.
Heterogeneity arises because traders have different distribution
assumptions about an informed trader's private signal, that is to
say traders agree to disagree with the precision of the signal.
Using the same description as Wang (1998), a trader is overconfident
if his distribution of the signal is too tight and underconfident if
it is too loose.

We give the existence and the uniqueness of the insider's
equilibrium trading strategy in a multi-period rational expectation
framework and give the analysis of the equilibrium. In equilibrium,
the ``trade public disclosure'' and ``heterogeneous prior beliefs''
have great effects on the insider's trading intensity, the market
depth and the effectiveness of the price.

We obtain many new and interesting results on market characteristics
and traders' strategies. First, under disclosure requirement, the
adverse selection decreases when there is an irrational informed
trader participating in the market, and the descending range is
positive related to the degree of heterogeneity. Since the insider
trades more aggressive than he should rationally do, he submits
larger orders. The market makers, realizing that the insider is
irrational and aggressive behavior, increase market depth. While
market makers decrease the market depth if there is no public
disclosure requirements, even though the insider is irrational. That
is to say, ``public disclosure'' leads to smaller market depth.

Also, under the disclosure requirement, the irrational insider's
private information is revealed slowly and linearly, for any degree
of his heterogeneity.  That is to say, under the  public disclosure
requirement, ``heterogeneous beliefs''  has no effect on the speed
of revelation of information. While ``heterogeneous beliefs'' has
great impacts on the trading behavior of irrational insiders and
market structure. In particular, under the disclosure requirement,
the insider dissembles his information by adding a random component
to his trades in every round except the last one. Despite this, our
analysis shows that the information is reflected more rapidly in
price with disclosure of insider trades than without.  An
interesting finding is that when the insider is overconfident or
underconfident, the trading intensity of the insider, the
heterogeneity parameter and the expected profits all fluctuate
greatly during all early auctions. We reveal the relationship
between the heterogeneity  and the trade intensity in the presence
of information disclosure, which agrees with our intuition in the
financial market.  That is to say, if the underconfident insider
puts a positive weight on asymmetric information this period, then
he puts a negative weight on heterogeneous prior beliefs this
period, and the case is inverse in the next period. We also show
that the co-existence of public disclosure and heterogeneous prior
beliefs leads to large and fluctuated trading volume and the
fluctuation is positively related to the degree of the insider's
heterogeneity.

More importantly, the irrational insider's  profits of some trading
rounds may be negative under the mandatory disclosure requirement.
The expected profits fluctuate during all auctions, and the
fluctuation which is positive related to the degree of heterogeneity
is small at the early auctions and becomes larger as the trading
goes by. Moreover, even though the irrational insider may get
negative profits in some periods, he trades to make sure the profit
of the last period and the whole trading profit are all positive.

This paper is structured as follows. In Section $2$ we introduce the
model and  in Section $3$ we make an analysis and give the
two-period equilibrium of the models with and without disclosure of
insider trades, respectively. In Section $4$, we give the unique
linear Nash equilibrium in multi-period framework and give the
analysis of the equilibrium.
 Finally,  the appendix contains the proof of some necessary propositions.

\section{The model}\setcounter{equation}{0}

\quad We conform to the notation of  Kyle (1985).  A single risky
asset with a terminal value
 $\tilde{v}$, which is
normally distributed with mean $p_0$ and variance $\Sigma_0$,
  is traded in an $N$-period sequential
auction market among three kinds of risk-neutral traders: a
monopolistic informed trader, noise traders and market makers. The
monopolistic informed trader  has a unique access to a private
signal $\tilde{s}*$  about $\tilde{v}$. All market makers and the
insider agree that the signal $\tilde{s}*$ is a scalar multiple of
the terminal value $\tilde{v}$, but they disagree concerning the
right scale, that is to say, the informed trader thinks
$\tilde{s}*=c_1\tilde{v} $ while the market makers think
$\tilde{s}*=c_2\tilde{v}$ , where $c_1$ and $c_2$ are different
positive constants. If we  consider a normalized signal
$\tilde{s}=\frac{\tilde{s}^*}{c_2}$, using the same analysis as Wang
(1998) we know that each trader's heterogeneous prior belief is
characterized completely by a parameter $K$,  where $K=c_2/c_1$ is a
positive constant. Then the informed trader thinks
$\tilde{s}=\tilde{v}/K$ while the market makers think
$\tilde{s}=\tilde{v}$.   If we consider the market maker's beliefs
(i.e., $K=1$) as the benchmark case, then the informed trader is
``overconfident'' if $ K>1$ and ``underconfident'' if $0<K<1$. The
noise traders (uninformed liquidity traders)  trade randomly; and
market makers set prices efficiently in the semi-strong form sense,
conditional on the total quantities traded by the informed trader
and noise traders, but not each of them. We also assume that there
is no discount across periods, i.e., the interest rate is normalized
to zero. This market model has some game-theoretic features. We can
view it as a game played by the insider and the market makers: the
insider attempts to hide his private information and make the best
use of his information to maximize his  profit; the market makers
attempt to learn the private information from the order and set
prices as efficiently as possible in order to rule out the profit
opportunities of the insider.

Let $\tilde{x}_n$ denote the market order submitted by the informed
trader at the $n$-th auction, conditional on his information, and
let $\tilde{\mu}_n$ denote  the random quantity traded by noise
traders at the $n$-th auction. We assume that $\tilde{\mu}_n$ is
normally distributed with zero mean and variance $\sigma_{\mu}^2,$
$n=1,2,\cdots, N,$ and
$\tilde{\mu}_1,\tilde{\mu}_2,\cdots,\tilde{\mu}_N, \tilde{v}$ are
mutually independent.
 So the total trading volume, denoted by $\tilde{y}_n$, is given by
 $\tilde{y}_n=\tilde{x}_n+\tilde{\mu}_n.$ The market makers set price
 $\tilde{p}_n$ in semi-strong form sense based on his information
 $\tilde{y}_1,\tilde{y}_2,\cdots,\tilde{y}_n$ at the $n$-th auction. Denote by $E_K[\cdots]$ and $E_1[\cdots]$ the expectation operators
of the informed trader and the market makers, respectively. Then
$\tilde{p}_n=E_1[\tilde{v}|\tilde{y}_1,\cdots, \tilde{y}_n]$,
$n=1,2,\cdots,N$.

 Let $\tilde{\pi}_n$ be the profit which accrues to the informed trader from
the $n$-th auction on, i.e.,  for $n=1,2\cdots,N,$
$\tilde{\pi}_n=\sum_{k=n}^N(\tilde{v}-\tilde{p}_k)\tilde{x}_k$. At
each auction the informed trader maximizes his total expected
profits of the current and the remaining rounds of trading
conditional on his information, i.e.,
$$\max_x E_K[\tilde{\pi}_n|\tilde{s}=s, \tilde{p}_1=p_1,\cdots,\tilde{p}_n=p_n],~~~~for ~~~n=1,2,\cdots,N.$$

The equilibrium conditions are that the competition between market
makers drives their expected profits to zero conditional on the
order flow and the fact that the  informed trader selects the
optimal strategy conditional on his correct conjectures and his
information at each auction. Following the convention in the
existing literature, an equilibrium is said to be linear if the
pricing rule is an affine function of the order flow. We will give
the definition of equilibrium of two-period and N-period in the
following sections, respectively.

\section{Analysis}
\setcounter{equation}{0}
\subsection{Two-period model without public disclosure of insider trades} \setcounter{equation}{0}

\quad\quad In order to get a benchmark against which to compare an
equilibrium for the case where the insider's trade in the first
period is publicly disclosed on completion of trading in that
period, we first give the two-period equilibrium of the model
without public disclosure of insider trades.

 The proposition below is based on a special case of Theorem $1$ of
Wang (1998).
\begin{proposition}\label{pro3.1}
 Given no public disclosure of insider trades, for $0<K<2,$\footnote{Just as the analysis in Wang (1998), the inequality
condition $0<K<2$ means that if a rational informed trader thinks a
risky asset is worth $100$, then an irrational informed trader's
subjective value cannot be less than $0$ or more than $200$ for the
equilibrium to exist.} a
 subgame perfect linear equilibrium exists. In this equilibrium
 there are constants
$\beta_n, \theta_n, \lambda_n, \gamma_n,
 \alpha_n, \omega_n, \phi_n, \delta_n$ and  $\Sigma_n ~(n=1,2)$ such
 that
$$\tilde{x}_n=\beta_n(1+\gamma_n)(\tilde{s}-\tilde{p}_{n-1})+\theta_n\tilde{s},$$
$$\tilde{p}_n-\tilde{p}_{n-1}=\gamma_n\tilde{p}_{n-1}+\lambda_n(\tilde{x}_n+\tilde{\mu}_n),$$
$$\Sigma_1=Var_1\{\tilde{v}|\tilde{x}_1+\tilde{\mu}_1\},$$
$$\Sigma_2=Var_1\{\tilde{v}|\tilde{x}_1+\tilde{\mu}_1,\tilde{x}_2+\tilde{\mu}_2\},$$
$$E_K(\tilde{\pi}_1|\tilde{s}=s,p_0)=\alpha_0(s-p_0)^2+\omega_0p_0s+\phi_0s^2+\delta_0,$$
$$E_K(\tilde{\pi}_2|\tilde{s}=s,\tilde{p}_1=p_1,p_0)=\alpha_1(s-p_1)^2+\omega_1p_1s+\phi_1s^2+\delta_1.$$
Given $\Sigma_0$ and $\sigma_{\mu}^2$, the above constants
$\beta_n,\theta_n,\lambda_n
,\gamma_n,\alpha_n,\omega_n,\phi_n,\delta_n$ and $\Sigma_n$$(n=1,2)$
satisfy:
$$\lambda_1=\frac{1}{\sigma_{\mu}}\frac{\sqrt{2K(2-K)[2-m(2-K)][1-m(1-K)]\Sigma_0}}{4-m(2-K)^2},~
\lambda_2=\frac{1}{2\sigma_{\mu}}\sqrt{\frac{2K(2-K)^2[1-m(1-K)]\Sigma_0}{4-m(2-K)^2}},$$
$$\beta_1=\frac{[2-m(2-K)^2]}{\lambda_1[4-m(2-K)^2]}, ~~~~\beta_2=\frac{1}{2\lambda_2},~~\theta_1=-\frac{\gamma_1}{\lambda_1},
~~\theta_2=\frac{K-1}{\lambda_2},
$$
$$\gamma_1=(1-K)[1-m(2-K)],~~~\gamma_2=1-K,$$
$$\alpha_0=\frac{(1+\gamma_1)^2}{4\lambda_1(1-\alpha_1\lambda_1)},~~~~\alpha_1=\frac{(2-K)^2}{4\lambda_2},$$
$$\omega_0=\omega_1+\frac{(2-K)\gamma_1}{\lambda_1},~~~\omega_1=\frac{(2-K)(1-K)}{\lambda_2},$$
$$\phi_0=\phi_1-\frac{\gamma_1}{\lambda_1},~~~\phi_1=\frac{K-1}{\lambda_2},$$
$$ \delta_0=\alpha_1\lambda_1^2\sigma_{\mu}^2, ~~~\delta_1=0,$$
$$\Sigma_1=\frac{2(2-K)[1-m(1-K)]}{4-m(2-K)^2}\Sigma_0,~
\Sigma_2=\frac{(2-K)^2[1-m(1-K)]}{4-m(2-K)^2}\Sigma_0$$ the boundary
condition is  $\alpha_2=\omega_2=\phi_2=\delta_2=0$, and the second
order condition is
 $\lambda_n(1-\alpha_n\lambda_n)>0,(n=1,2)$, where $m\doteq
\frac{\lambda_1}{\lambda_2}$, and $m$ is the unique solution of the
following equation
$$(2-K)^3m^3-4(2-K)m^2-4(2-K)m+8=0.$$
satisfying  $0<m<\frac{2}{2-K}$.
\end{proposition}

\begin{proof}See Appendix.
\end{proof}

\subsection{Two-period model with public
disclosure of
          insider trades} \setcounter{equation}{0}

\quad\quad Using  the same method  by Huddart, Hughes and Levine
(2001), we know that no invertible trading strategy
 can be part of an equilibrium in this case, and
 we  show that there exists  an equilibrium in which the
insider's first-period trade consists of an information-based linear
component
$\beta_{1}(1+\gamma_1)(\tilde{s}-\tilde{p}_{0})+\theta_1\tilde{s}$
and a noise component $\tilde{z}_1$, which is independently of
$\tilde{v}$ and $\tilde{\mu}_1$ and normally distributed with mean
$0$ and variance $\sigma_{z_1}^2$. For
 market makers, the public disclosure of $x_1$ allows them  to
update their beliefs based on  the first period order flow. In
particular, let $\tilde{p}_1^*=p_0+\gamma_1'p_0+\eta_1\tilde{x}_1$
be the expected value of $\tilde{v}$ given $\tilde{x}_1=x_1$ and
$\tilde{y}_1=y_1$. Thus,
$E_1(\tilde{v}|\tilde{x}_1=x_1,\tilde{y}_1=y_1)=E_1(v|\tilde{x}_1=x_1)$.
In turn, $\tilde{p}_1^*$ replaces $\tilde{p}_1$ in the second period
price
$\tilde{p}_2=\tilde{p}_1^*+\gamma_2\tilde{p}_1^*+\lambda_2\tilde{y}_2=\tilde{p}_1^*+\gamma_2\tilde{p}_1^*+\lambda_2(\tilde{x}_2+\tilde{\mu}_2)$.

 Applying the principal of
backward induction, we can write the insider's second period
optimization problem for given $\tilde{x}_1=x_1$ and
$\tilde{p}_1^*=p_1^*$ as  $x_2\in
\arg\max_xE_K[x(\tilde{v}-\tilde{p}_2)|\tilde{x}_1=x_1,\tilde{p_1^*}=p_1^*,\tilde{s}=s].$
Since
\begin{equation}{\label{eq3.1}}
\begin{aligned}
\max_xE_K[\tilde{x}(\tilde{v}-\tilde{p}_2)|\tilde{x}_1=x_1,\tilde{p_1^*}=p_1^*,\tilde{s}=s]
=\max_x[(Ks-p_1^*-\gamma_2p_1^*-\lambda_2x)x],
\end{aligned}
\end{equation}
the first order condition implies
\begin{equation}\label{eq3.3}
x_2=\beta_2(1+\gamma_2)(s-{p}_1^*)+\theta_2{s}=\frac{1+\gamma_2}{2\lambda_2}(s-p_1^*)+\frac{K-1-\gamma_2}{2\lambda_2}s,
\end{equation}
and the second order condition is $\lambda_2>0$. So
\begin{equation}\label{eq3.4}
E_K[\tilde{\pi}_2(\tilde{p}_1^*,\tilde{s})|\tilde{x}_1=x_1,\tilde{p_1^*}=p_1^*,
\tilde{s}=s]=\frac{1}{4\lambda_2}[Ks-(1+\gamma_2)p_1^*]^2,
\end{equation}
and
\begin{equation}\label{eq3.5}
\beta_2=\frac{1 }{2\lambda_2},~~~
\theta_2=\frac{K-1-\gamma_2}{2\lambda_2}.
\end{equation}
Substituting the above into
$\tilde{p}_2=\tilde{p}_1^*+\gamma_2\tilde{p}_1^*+\lambda_2\tilde{y}_2=\tilde{p}_1^*+\gamma_2\tilde{p}_1^*+\lambda_2(\tilde{x}_2+\tilde{\mu}_2)$,
and using $E(\tilde{p}_2^*-\tilde{p}_1^*)=0$, we obtain
\begin{equation}\label{eq3.5'}
\gamma_2=1-K.
\end{equation}
Stepping back to the insider's first period optimization problem, we
have
$$ x_1\in
 \arg\max_xE_K[\tilde{x}(\tilde{v}-\tilde{p}_1)+\pi_2(\tilde{p}_1^*,\tilde{s})|\tilde{s}=s],
$$
and
\begin{equation}\label{eq3.6}\begin{aligned}
E_K[\tilde{x}(\tilde{v}-\tilde{p}_1)+\tilde{\pi}_2(\tilde{p}_1^*,\tilde{s})|\tilde{s}=s]
=x[Ks-p_0-\gamma_1p_0-\lambda_1x]+\frac{1}{4\lambda_2}[Ks-(1+\gamma_2)(p_0+\gamma_1'p_0+\eta_1x)]^2.
\end{aligned}
\end{equation}
The first order condition implies
\begin{equation}{\label{eq3.8}}\begin{aligned}
&\left(\frac{\eta_1^2(1+\gamma_2)^2}{2\lambda_2}-2\lambda_1\right)x_1+
\left(K-\frac{\eta_1(1+\gamma_2)K}{2\lambda_2}\right)s\\-
&\left[(1+\gamma_1)-\frac{\eta_1(1+\gamma_2)^2(1+\gamma_1')}{2\lambda_2}\right]p_0=0.
\end{aligned}
\end{equation}
If our proposed mixed trading strategy
 $\tilde{x}_{1}=\beta_{1}(1+\gamma_1)(\tilde{s}-\tilde{p}_{0})+\theta_1\tilde{s}+\tilde{z}_1$
is to hold in the equilibrium,  we can seek values of $\lambda_1$,
$\lambda_2$, and $\gamma_1$  from Eq. $(\ref{eq3.8})$ such that
$\lambda_1>0,\lambda_2>0$, and
\begin{equation}{\label{eq3.9}}\begin{aligned}
\frac{\eta_1^2(1+\gamma_2)^2}{2\lambda_2}-2\lambda_1=0,
\end{aligned}
\end{equation}
\begin{equation}{\label{eq3.10}}\begin{aligned}
K-\frac{\eta_1(1+\gamma_2)K}{2\lambda_2}=0,
\end{aligned}
\end{equation}
\begin{equation}{\label{eq3.11}}\begin{aligned}
(1+\gamma_1)-\frac{\eta_1(1+\gamma_2)^2(1+\gamma_1')}{2\lambda_2}=0.
\end{aligned}
\end{equation}
Combining Eqs. $(\ref{eq3.9})$, $(\ref{eq3.10})$ with
$(\ref{eq3.11})$, we  get
$$\lambda_1=\lambda_2=\frac{\eta_1(1+\gamma_2)}{2}.$$
By the market's  efficient condition, we have
\begin{equation}{\label{eq3.13}}\tilde{p}_1=E_1(\tilde{v}|\tilde{x}_1+\tilde{\mu}_1)=p_0+\gamma_1p_0+
\lambda_1(\tilde{x}_1+\tilde{\mu}_1),~~~\tilde{p}_1^*=E_1(\tilde{v}|\tilde{x}_1,\tilde{y}_1)=p_0+\gamma_1'p_0+\eta_1\tilde{x}_1,\end{equation}
where
\begin{equation}{\label{eq3.16}}\lambda_1=\frac{[\beta_1(1+\gamma_1)+\theta_1]\Sigma_0}
{[\beta_1(1+\gamma_1)+\theta_1]^2\Sigma_0+\sigma_{\mu}^2+\sigma_{z_1}^2},
\end{equation}
\begin{equation}{\label{eq3.18}}\eta_1=\frac{[\beta_1(1+\gamma_1)+\theta_1]\Sigma_0}
{[\beta_1(1+\gamma_1)+\theta_1]^2\Sigma_0+\sigma_{z_1}^2},
\end{equation}
\begin{equation}{\label{eq3.17}}\begin{aligned}\gamma_1=-\lambda_1\theta_1,
~~~\gamma_1'=-\eta_1\theta_1.
\end{aligned}
\end{equation}
Note that
$$\tilde{p}_2-\tilde{p}_1^*=\gamma_2\tilde{p}_1^*+\lambda_2\tilde{y}_2=\gamma_2\tilde{p}_1^*+\lambda_2(\tilde{x}_2+\tilde{\mu}_2)$$
and  $$\tilde{p}_2-\tilde{p}_1^*=E_1(\tilde{v}-
\tilde{p}_1^*|\tilde{x}_2+\tilde{\mu}_2,\tilde{p}_1^*), $$ by the
projection theorem, we know that $\tilde{p}_2-\tilde{p}_1^*$ is the
projection of $\tilde{v}- \tilde{p}_1^*$ onto the two dimension
space spanned by $(\tilde{x}_2+\tilde{\mu}_2,\tilde{p}_1^*)$ in
which $\lambda_2$ is the coefficient of the $2$ed orthogonal basis
$\frac{\gamma_2}{\lambda_2}\tilde{p}_1^*+\tilde{x}_2+\tilde{\mu}_2.$
Thus, we obtain
\begin{equation}{\label{eq3.20}} \begin{aligned}\tilde{p}_2-\tilde{p}_1^*=&E_1(\tilde{v}-
\tilde{p}_1^*|\tilde{x}_2+\tilde{\mu}_2,\tilde{p}_1^*)=
E_1\left[\tilde{v}-\tilde{p}_1^*|\frac{\gamma_2}{\lambda_2}\tilde{p}_1^*+\tilde{x}_2+\tilde{\mu}_2\right]\\
=&E_1\left[\tilde{v}-\tilde{p}_1^*|(\frac{1+\gamma_2}{2\lambda_2}-\frac{\gamma_2}{\lambda_2})
(\tilde{s}-p_1^*)+\tilde{\mu}_2\right].
\end{aligned}
\end{equation}
Furthermore,
\begin{equation}\label{eq3.21}\lambda_2=\frac{\frac{1-\gamma_2}{2\lambda_2}\Sigma_1}
{\frac{(1-\gamma_2)^2}{4\lambda_2^2}\Sigma_1+\sigma_{\mu}^2},
~~~\gamma_2=1-K.
\end{equation}
Then it is easy to get
\begin{equation}\label{eq3.23}\lambda_2=\frac{1}
{2\sigma_{\mu}}\sqrt{(2-K)K\Sigma_1} ,
\end{equation}
where $K$ must satisfy $0<K<2.$

Next, we consider the effectiveness of the price, measured by
$\Sigma_n (n=1,2).$ Note that
\begin{equation}\label{eq3.33}\begin{aligned}
\Sigma_1=Var(\tilde{v}|\tilde{p}_1^*)=Var(\tilde{v}|\tilde{x}_1)
=\Sigma_0-\eta_1^2[(\beta_1(1+\gamma_1)+\theta_1)^2\Sigma_0+\sigma_{z_1}^2].
\end{aligned}
\end{equation}
Using $\lambda_1=\lambda_2=\frac{\eta_1(2-K)}{2}$, we obtain
\begin{equation}\label{eq3.34}\begin{aligned}
\frac{(\beta_1(1+\gamma_1)+\theta_1)\Sigma_0}{(\beta_1(1+\gamma_1)+\theta_1)^2\Sigma_0+\sigma_{z_1}^2+\sigma_{\mu}^2}
=\frac{2-K}{2}\frac{(\beta_1(1+\gamma_1)+\theta_1)\Sigma_0}{(\beta_1(1+\gamma_1)+\theta_1)^2\Sigma_0+\sigma_{z_1}^2},
\end{aligned}
\end{equation}
i.e.,
\begin{equation}\label{eq3.35}\begin{aligned}
K[(\beta_1(1+\gamma_1)+\theta_1)^2\Sigma_0+\sigma_{z_1}^2+\sigma_{\mu}^2]=2\sigma_{\mu}^2.
\end{aligned}
\end{equation}
By Eq. $(\ref{eq3.11})$ we get
$$(1-\lambda_1\theta_1)-(1-\eta_1\theta_1)(2-K)=0,$$
i.e.,
$$\theta_1=\frac{1-K}{\lambda_1}.$$
Then Eq.(\ref{eq3.16}) can be rewritten as
\begin{equation}\label{eq3.36}\lambda_1=\frac{K\beta_1(1+\gamma_1)\Sigma_0+\frac{K(1-K)}{\lambda_1}\Sigma_0}{2\sigma_{\mu}^2},\end{equation}
and thus
\begin{equation}\label{eq3.37}\begin{aligned}
\beta_1(1+\gamma_1)=\frac{2\sigma_{\mu}^2\lambda_1}{K\Sigma_0}-\frac{(1-K)}{\lambda_1}.
\end{aligned}
\end{equation}
Eqs. $(\ref{eq3.23})$, $(\ref{eq3.33})$  and $(\ref{eq3.35})$ yield
\begin{equation}\label{eq3.38}\begin{aligned}
\Sigma_1=\Sigma_0-\frac{4\lambda_2^2}{K(2-K)}\sigma_{\mu}^2=\Sigma_0-\Sigma_1,
\end{aligned}
\end{equation}
i.e.,
$$\Sigma_1=\frac{1}{2}\Sigma_0.$$
From the above we give the following proposition:

\begin{proposition}\label{pro3.2}For $0<K<2$ , there exist a unique linear equilibrium in the
two-period setting with public disclosure of the irrational
insider's trades, in which there are constants
$\beta_n,\theta_n,\lambda_n,\gamma_n ~~(n=1,2)$,  $\gamma_1',
\eta_1$ and  $\Sigma_1$  such that:
$$\tilde{x}_1=\beta_1(1+\gamma_1)(\tilde{s}-\tilde{p}_0)+\theta_1\tilde{s}+\tilde{z}_1,~~~
 \tilde{x}_2=\beta_2(1+\gamma_2)(\tilde{s}-\tilde{p}_1^*)+\theta_2\tilde{s},$$
$$\tilde{p}_1-p_0=\gamma_1p_0+\lambda_1(\tilde{x}_1+\tilde{\mu}_1),~~~
 \tilde{p}_1^*-p_0=\gamma_1'p_0+\eta_1\tilde{x}_1,$$
$$\tilde{p}_2-\tilde{p}_1^*=\gamma_2\tilde{p}_1^*+\lambda_2(\tilde{x}_2+\tilde{\mu}_2)
,~~\Sigma_1=Var_1\{\tilde{v}|\tilde{x}_1\}.$$
Given $\Sigma_0$ and
$\sigma_{\mu}^2$, the constants $\beta_n,\theta_n,\lambda_n,\gamma_n
~~(n=1,2)$,  $\gamma_1', \eta_1$ and  $\Sigma_1$ are characterized
as follows:
$$\lambda_1=\lambda_2=\frac{\sqrt{K(2-K)\Sigma_0}}{2\sqrt{2}\sigma_{\mu}},$$
$$\eta_1=\frac{\sqrt{K\Sigma_0}}{\sigma_{\mu}\sqrt{2(2-K)}},$$
$$\gamma_1=K-1,~~\gamma_1'= \frac{2(K-1)}{2-K},~~\gamma_2=1-K,$$
$$\beta_1=\frac{(3K-2)\sigma_{\mu}}{K\sqrt{2K(2-K)\Sigma_0}},~~~~~
 \beta_2=\sigma_{\mu}\sqrt{\frac{2}{(2-K)K\Sigma_0}};$$
$$\theta_1=\frac{2\sqrt{2}(1-K)\sigma_{\mu}}{\sqrt{K(2-K)\Sigma_0}},~~~~~
\theta_2 =-\theta_1;$$
$$E(\tilde{\pi}_1)=\frac{K^2}{\sqrt{2K(2-K)}}\sigma_{\mu}\sqrt{\Sigma_0}+\frac{6K^2-10K+4}{\sqrt{2K(2-K)}}\frac{\sigma_{\mu}}{\sqrt{\Sigma_0}}p_0^2,,$$
$$E(\tilde{\pi}_2)=\frac{5K^2-8K+4}{2\sqrt{2K(2-K)}}\sigma_{\mu}\sqrt{\Sigma_0}+\frac{(2K-2)^2}{\sqrt{2K(2-K)}}\frac{\sigma_{\mu}}{\sqrt{\Sigma_0}}p_0^2,,$$
$$\Sigma_1=\frac{1}{2}\Sigma_0,$$
$$\sigma_{z_1}^2=\frac{2-K}{2K}\sigma_{\mu}^2.$$
\end{proposition}

\begin{proof} See  Appendix.
\end{proof}

{\bf Analysis of the equilibrium}: The parameters $\beta_n$ and
$\theta_n$ measure the intensity of trading due to asymmetric
information and   heterogeneous prior beliefs, respectively.  In
Proposition \ref{pro3.2}, we can have
$\frac{\beta_{1}}{\beta_{2}}=\frac{3K-2}{2K}<1$, which implies the
intensity of trading on asymmetric information at the first period
is lower than that at the second period ($\beta_1<\beta_2$), no
matter what the insider's heterogeneous prior belief is.  These are
consistent with the absence of a concern for the effect of trading
on future expected profits in the last period.   It also implies
that when the insider is underconfident ($0<K<1$), he puts a
positive weight on heterogeneous prior beliefs in the first period
and negative weight in the second period, i.e., $\theta_1>0$ and
$\theta_2=-\theta_1<0$. While the insider is overconfident, he does
the opposite, that is to say, he puts a negative weight on
heterogeneous prior beliefs in the first period and positive weight
in the second period, i.e., $\theta_1<0$ and $\theta_2=-\theta_1>0$.
These are consistent with our intuition.

The parameters $\lambda_n$, $\gamma_n$ and $\gamma_n'$ characterize
the pricing rule. The liquidity parameter $\lambda_n$ is an inverse
measure of market depth and the heterogeneous parameter $\gamma_n$
measures the correction in the efficient price change per unit of
the price in the previous period due to heterogeneous prior beliefs.
The   parameter $\gamma_n'$ measures the adjusted correction. From
the  proposition above, we know that the market liquidity are the
same across the two periods, i.e. $\lambda_1=\lambda_2$, no matter
what the degree of the heterogenous belief is.
$\eta_1-\lambda_1=\frac{K\sqrt{K\Sigma_0}}{2\sigma_{\mu}\sqrt{2(2-K)}}$
equals the price adjustment based on the first period order flow and
it increasing with $K$.

 The measure of the informativeness of price, $\Sigma_1$, equals
 $\frac{\Sigma_0}{2}$. That is to say, at the end of the first
 trading period, a half of the private information has been
 incorporated into the price. Setting $\sigma_{z_1}^2=\frac{2-K}{2K}\sigma_{\mu}^2$
 serves to disguise the information based component of the insider's  trades once
 they are publicly disclosed. It is easy to see that
 $\sigma_{z_1}^2$ is a decreasing function of the parameter $K$,
 That is to say the more confident of the insider the less noise he
 puts in formulating his strategy.

 It is worth noting that the insider's  expected profit at each period   is
 not always positive as the existence literatures. The sign of the
 expected profit depends on the degree of heterogeneity. For
 example, the insider's first period expected profit is
 $E(\tilde{\pi}_1)-E(\tilde{\pi}_2)=\frac{-3K^2+8K-4}{2\sqrt{2K(2-K)}}
 \sigma_{\mu}\sqrt{\Sigma_0}+\frac{2K^2-2K}{\sqrt{2K(2-K)}}\frac{\sigma_{\mu}}{\sqrt{\Sigma_0}}p_0^2$,
 and when $0<K<\frac{2}{3}$, $E(\tilde{\pi}_1)-E(\tilde{\pi}_2)<0$.

The same exogenous parameters imply different values for the
endogenous parameters  depending  not only on whether the insider
must disclose  his trader after the fact or not but also on the
degree of heterogeneity. In order to distinguish these parameters,
we add an upper bar to the endogenous parameters in the case of
Huddart, Hughes and Levine (2001)'s model and a hat to the
endogenous parameters in the case of Wang (1998). Note that when
$K=1$ the above proposition  is just Proposition $2$ of Huddart,
Hughes and Levine (2001). The next proposition compares the
endogenous parameters across Huddart, Hughes and Levine (2001)'s
model and  our model when $K\neq 1.$

\begin{proposition}\label{pro3.3}In the two-period setting, the endogenous
parameters  across Huddart, Hughes and Levine (2001)'s model and
our model satisfy:

(i) when $0<K<1$,
  $$\beta_1(1+\gamma_1)+\theta_1>\bar{\beta}_1,
 ~~~\beta_2(1+\gamma_2)+\theta_2<\bar{\beta}_2,$$
when $1<K<2$,
$$\beta_1(1+\gamma_1)+\theta_1<\bar{\beta}_1, ~~~\beta_2(1+\gamma_2)+\theta_2>\bar{\beta}_2;$$

(ii) for all $0<K<2, K\neq 1$,
$$\lambda_1=\lambda_2<\bar{\lambda}_1=\bar{\lambda}_2;$$

(iii) for all  $0<K<2, K\neq 1$,
$$ \Sigma_1=\bar{\Sigma}_1=\frac{\Sigma_0}{2};$$

(iv) when $0<K<\frac{2}{3}$, $E(\pi_1)$ is bigger or smaller than
$\frac{\sigma_{\mu}\sqrt{\Sigma_0}}{\sqrt{2}}$ (the total profit of
two period in the Huddart, et al. (2001)'s model)  depends on the
value of $p_0$. Especially,
 when $\frac{2}{3}<K<1$,
$E(\pi_1)<\frac{\sigma_{\mu}\sqrt{\Sigma_0}}{\sqrt{2}}$, and  when
$1<K<{2}$, $E(\pi_1)>\frac{\sigma_{\mu}\sqrt{\Sigma_0}}{\sqrt{2}}$.

(v) When $0<K<1$, $\sigma_{z_1}^2>\bar{\sigma}_{z_1}^2$,
 and when $1<K<2$, $\sigma_{z_1}^2<\bar{\sigma}_{z_1}^2.$

\end{proposition}
\proof: See the Appendix.

 In our model, $\beta_i(1+\gamma_i)+\theta_i$ ($i=1,2$)
characterize the degree of the  dependence of the private
information, from the Proposition \ref{pro3.3} we know that under
the requirement of trade disclosure,
 the underconfident insider's first period
trading intensity is bigger than the rational case, while the
overconfident insider's  is smaller than the rational case. ``Trade
disclosure'' makes the overconfident trader trade less aggressively
than the underconfident trader in the first period. But in the
second period, the trading intensity of the overconfident insider is
bigger than the rational and the underconfident trader's trading
intensity. Not
 surprisingly, when the insider is irrational, whatever overconfident or underconfident, the
 ``depth''
of the market, i.e. the order flow
 necessary to induce prices to rise or fall by one dollar, is
 bigger  than that of Huddart, et al. (2001)'s.

The measure of the informativeness of price $\Sigma_1$ equals
$\frac{\Sigma_0}{2}$. This means that in the two-period model ``the
public disclosure'' and ``heterogeneous prior beliefs'' have no
effect on the informativeness of the price, but it is not true when
the trading period is bigger than two, which will be discussed in
Section $4$.

In order to disguise his trading, the insider puts a noise in
formulating his strategy whose variance is
$\frac{2-K}{2K}\sigma_{\mu}^2$. From $(v)$ in Proposition
\ref{pro3.3}, we know that when the insider is underconfident the
variance of the noise on his trading strategy  is bigger than the
rational trader, and when he is overconfident  the variance of the
noise on his trading strategy is smaller than the rational case.
This coincides with our intuition.

The following proposition compares the endogenous parameters across
the case of our model and Wang (1998)'s model. For convenience, we
only analysis the cases of $K= 0.5$ and  $K=1.5$, respectively.

\begin{proposition}\label{pro3.4} In the two-period setting, the endogenous parameters across
the case of our model and Wang (1998)'s model satisfy:

(i) When $K=0.5$,
$$\lambda_1=\lambda_2<\hat{\lambda}_1<\hat{\lambda}_2,~~\beta_1<\hat{\beta}_1<0,
~~\beta_2>\hat{\beta}_2>0,~~\Sigma_1<\hat{\Sigma}_1,$$
$$\gamma_1<\hat{\gamma}_1<0,~~\gamma_2=\hat{\gamma}_2,~~\theta_1>\hat{\theta}_1>0, ~~\theta_2<\hat{\theta}_2<0,$$
$$E(\pi_1-\pi_2)<0<E(\hat{\pi}_1-\hat{\pi}_2),~~E(\pi_2)>E(\hat{\pi}_2),$$
(ii) When $K=1.5$,
$$\lambda_1=\lambda_2<\hat{\lambda}_2<\hat{\lambda}_1,~~\beta_1>\hat{\beta}_1>0,
~~\beta_2>\hat{\beta}_2>0,~~\Sigma_1<\hat{\Sigma}_1,$$
$$\gamma_1>0>\hat{\gamma}_1,~~\gamma_2=\hat{\gamma}_2,~~\theta_1<0<\hat{\theta}_1, ~~\theta_2>\hat{\theta}_2>0.$$
$E(\pi_1-\pi_2)$ is bigger or smaller than
$E(\hat{\pi}_1-\hat{\pi}_2)$, depending on the values of $p_0$ and
$\Sigma_0$. So do  $E(\pi_2)$ and $E(\hat{\pi}_2)$.
\end{proposition}

The above proposition implies that ``public disclosure requirement''
leads to more effectiveness  of price and larger market depth,
whenever the insider is overconfident or underconfident.
 If the insider is underconfident, he puts a smaller weight on his
 private information in the first period and a larger in the second
 period. However, the overconfident insider puts a larger weights in
 the two periods under the disclosure requirement. The proposition
 also implies even though ``public disclosure requirement'' may lead to
 negative profit in the first period the irrational insider can
 benefit from whole insider trading.

From the discussion above we can conclude that ``the public
disclosure'' and ``heterogeneous prior beliefs'' have a great
influence on the equilibrium.

\section{A sequential auction equilibrium}

\quad\quad In this section we generalize the model into $N$-period
trading where a number of rounds of trading with public disclosure
take place sequentially. The model is structured such that the
equilibrium price at each auction reflects the information contained
in the past and the current order flow, and the  insider maximizes
his expected profits in the equilibrium taking into account his
effect on price in both the current auction and  the future auction.
\subsection{The sequential equilibrium}
\quad\quad Now,  we  represent a proposition which provides a
difference equation
 system to characterize the  equilibrium.

\begin{proposition}\label{pro4.1}In the economy with one irrational  informed
traders , for $0<K<2$,  there exist a unique subgame perfect
equilibrium and the equilibrium is a recursive equilibrium. In this
equilibrium there are constants
$\beta_n,\theta_n,\lambda_n,\gamma_n,\gamma_n',\alpha_n,\eta_n,\omega_n,\phi_n,\delta_n$
and $\Sigma_n$, characterized by the following:

\begin{equation}\label{eq4.1}\tilde{x}_n=\beta_n(1+\gamma_n)(\tilde{s}-p_{n-1}^*)+\theta_n\tilde{s}+\tilde{z}_n,\end{equation}
\begin{equation}\label{eq4.2}\tilde{p}_n-\tilde{p}_{n-1}^*=\gamma_n\tilde{p}_{n-1}^*+\lambda_n(\tilde{x}_n+\tilde{\mu}_n),\end{equation}
\begin{equation}\label{eq4.3}\tilde{p}_n^*-\tilde{p}_{n-1}^*=\gamma_n'\tilde{p}_{n-1}^*+\eta_n\tilde{x}_n,\end{equation}
\begin{equation}\label{eq4.4}\begin{aligned}&E_K[\tilde{\pi}_{n+1}|\tilde{p}_1^*=p_1^*,\cdots,\tilde{p}_{n}^*=p_{n}^*,
\tilde{p}_1=p_1,\cdots,\tilde{p}_{n}=p_n,\tilde{s}=s]\\&=\alpha_{n}(s-p_{n}^*)^2
+\omega_{n}sp_{n}^*+\phi_{n}s^2+\delta_{n},\end{aligned}\end{equation}
\begin{equation}\label{eq4.5}\Sigma_n=Var_1\{\tilde{v}|\tilde{p}_1^*,\cdots,\tilde{p}_{n}^*,\tilde{p}_1,\cdots,\tilde{p}_{n}\}.\end{equation}
Given $\Sigma_0$ and $\sigma_{\mu}^2$,
$\beta_n,\theta_n,\lambda_n,\gamma_n,\gamma_n',\alpha_n,\eta_n,
\omega_n,\phi_n,\delta_n$ and $\Sigma_n$ $(n=1,\cdots,N-1)$ are the
unique solution to the difference equation system
\begin{equation}\label{eq4.5}\begin{aligned}
\alpha_{n-1}=[1+\gamma_n-\lambda_n\beta_n(1+\gamma_n)]\beta_n(1+\gamma_n)+
\alpha_n(1+\gamma_n'-\eta_n\beta_n(1+\gamma_n))^2,
\end{aligned}
\end{equation}
\begin{equation}\label{eq4.6}\begin{aligned}
\omega_{n-1}=&\beta_n(1+\gamma_n)(1-K)-(1+\gamma_n-\lambda_n\beta_n(1+\gamma_n))\theta_n+
\omega_n[1-(\eta_n\beta_n(1+\gamma_n)-\gamma_n')],
\end{aligned}
\end{equation}
\begin{equation}\label{eq4.7}\begin{aligned}
\phi_{n-1}=\phi_n+\beta_n(1+\gamma_n)(K-1)+[K+\gamma_n-\lambda_n\beta_n(1+\gamma_n)]\theta_n+
\omega_n[\eta_n\beta_n(1+\gamma_n)-\gamma_n'],
\end{aligned}
\end{equation}
\begin{equation}\label{eq4.8}\begin{aligned}
\delta_{n-1}=\delta_n,
\end{aligned}
\end{equation}
\begin{equation}\label{eq4.9}\lambda_n=\frac{K^2\alpha_n}
{(2\alpha_n-\omega_n)^2},
\end{equation}
\begin{equation}\label{eq4.10}\eta_n=\frac{K}
{2\alpha_n-\omega_n},
\end{equation}
\begin{equation}\label{eq4.11}\theta_n=\frac{2\alpha_nK-2\alpha_n+\omega_n}
{(2\alpha_n-\omega_n)\lambda_n},
\end{equation}
\begin{equation}\label{eq4.12}
\beta_n(1+\gamma_n)+\theta_n=\frac{(2\alpha_n-\omega_n)\sigma_{\mu}^2\lambda_n}{(2\alpha_n-K\alpha_n-\omega_n)\Sigma_{n-1}},
\end{equation}
\begin{equation}\label{eq4.12'}
\gamma_n=-\lambda_n\theta_n, ~~~~\gamma'_n=-\eta_n\theta_n,
\end{equation}
\begin{equation}\label{eq4.13}\begin{aligned}
&\sigma_{z_n}^2=\frac{K\alpha_n\sigma_{\mu}^2}{2\alpha_n-K\alpha_n-\omega_n}-
\frac{(2\alpha_n-\omega_n)^2\sigma_{\mu}^4\lambda_n^2}{(2\alpha_n-K\alpha_n-\omega_n)^2\Sigma_{n-1}},
\end{aligned}
\end{equation}
\begin{equation}\label{eq4.14}\begin{aligned}\Sigma_n
=\Sigma_{n-1}-\frac{K\lambda_n}{2\alpha_n-K\alpha_n-\omega_n}\sigma_{\mu}^2,
\end{aligned}
\end{equation}
for all auction  $n=1,\cdots,N-1$, and
\begin{equation}\label{eq4.13'}\delta_{N-1}=0,~
\alpha_{N-1}=\frac{(2-K)^2}{4\lambda_N},~
\omega_{N-1}=\frac{(2-K)(1-K)}{\lambda_N},~
\phi_{N-1}=\frac{K-1}{\lambda_N},
\end{equation}
\begin{equation}\label{eq4.14'}\beta_{N}=\frac{1}{2\lambda_N},
~\lambda_{N}=\frac{\sqrt{K(2-K)\Sigma_{N-1}}}{2\sigma_{\mu}}, ~
\theta_N=\frac{K-1}{\lambda_N}, ~\gamma_N=1-K,
\end{equation}
 $\alpha_N=\omega_N=\phi_N=\delta_N=0.$
\end{proposition}
\begin{proof} See the Appendix.
\end{proof}

As in Wang (1998), the inequality condition for the belief parameter
$K$, i.e., $0<K<2$, is required in the above proposition. It results
from the fact that the second order condition, which is given by Eq.
(\ref{eqa4c}), should be satisfied.

 We will give the analysis of the
equilibrium described by the proposition above in the following.

\subsection{Equilibrium volume}

\quad\quad Given the equilibrium of our model, we now investigate
how ``public disclosure'' and ``heterogeneous prior beliefs'' affect
the behavior of the  trading volume.

 Following Admati and Pfleiderer (1988),  the total trading volume at the $n$ th
auction denoted by $Vol_n$, is defined by
\begin{equation}\label{eq4.15}Vol_n=\frac{1}{2}(|\tilde{x}_n|+|\tilde{y}_n|+|\tilde{\mu}_n|).
\end{equation}

Using the expressions of $\tilde{x}_n$, $\tilde{y}_n$ and
$\tilde{\mu}_n$ in Proposition \ref{pro4.1}, we can get the total
expected trading volume as shown in the following proposition:

\begin{proposition}\label{pro4.2}
Both ``public disclosure'' and  ``heterogeneous prior beliefs'' lead
to a larger trading volume and the expected trading volume at the
$n$th ($i=1,2,\cdots, N.$) auction is
\begin{equation}\label{eq4.16}E_1[Vol_n]=\frac{1}{\sqrt{2\pi}}(V_n^i+V_n^l+V_n^m).
\end{equation}
where
\begin{equation}\begin{aligned}\label{eq4.17}
V_n^i=&\sqrt{var_1(\tilde{x}_n)} =\\&
\sqrt{\left[\frac{(2\alpha_n-\omega_n)\sigma_{\mu}^2\lambda_n}{(2\alpha_n-K\alpha_n-\omega_n)
\Sigma_{n-1}}\right]^2\Sigma_{n-1}+
\left[\frac{2\alpha_nK-2\alpha_n+\omega_n}
{(2\alpha_n-\omega_n)\lambda_n}\right]^2(\Sigma_0+p_0^2-\Sigma_{n-1})+\sigma_{z_n}^2},
\end{aligned}
\end{equation}
\begin{equation}\begin{aligned}\label{eq4.19}V_n^m=&\sqrt{var_1(\tilde{y}_n)}\\
=&\sqrt{\left[\frac{(2\alpha_n-\omega_n)\sigma_{\mu}^2\lambda_n}{(2\alpha_n-K\alpha_n-\omega_n)
\Sigma_{n-1}}\right]^2\Sigma_{n-1}+
\left[\frac{2\alpha_nK-2\alpha_n+\omega_n}
{(2\alpha_n-\omega_n)\lambda_n}\right]^2(\Sigma_0+p_0^2-\Sigma_{n-1})+\sigma_{z_n}^2+\sigma_{\mu}^2}.
\end{aligned}
\end{equation}
\begin{equation}\label{eq4.18}V_n^l=\sqrt{var_1(\tilde{\mu}_n)}=\sigma_{\mu}
\end{equation}

for $n=1,2,\cdots, N-1.$ And
\begin{equation}\label{eq4.20}V_N^i=\sqrt{var_1(\tilde{x}_N)}=\sqrt{\frac{K\sigma_{\mu}^2}{(2-K)}
+\frac{4(K-1)^2\sigma_{\mu}^2}
{K(2-K)\Sigma_{N-1}}(\Sigma_0+p_0^2-\Sigma_{N-1})},
\end{equation}
\begin{equation}\label{eq4.22}V_N^m=\sqrt{var_1(\tilde{y}_N)}
=\sqrt{\frac{K\sigma_{\mu}^2}{(2-K)}+\frac{4(K-1)^2\sigma_{\mu}^2}
{K(2-K)\Sigma_{N-1}}(\Sigma_0+p_0^2-\Sigma_{N-1})+\sigma_{\mu}^2}.
\end{equation}
\begin{equation}\label{eq4.21}V_N^l=\sqrt{var_1(\tilde{\mu}_N)}=\sigma_{\mu},
\end{equation}
\end{proposition}
\begin{proof} See the Appendix.
\end{proof}

Proposition \ref{pro4.2} gives the contribution of each group of
traders to the total trading volume. From the proof of the
proposition \ref{pro4.1}, we know that $\omega_n=0$ when $K=1$. Then
the term  $\frac{2\alpha_nK-2\alpha_n+\omega_n}
{(2\alpha_n-\omega_n)\lambda_n}$ in equation (\ref{eq4.19}) and
(\ref{eq4.17})  vanishes when $K=1.$ That is to say both the insider
and the market makers trade a larger volume under heterogeneous
prior beliefs and thus lead to an increase in total trading volume.
Furthermore, $\sigma_{z_n}^2$ in equation (\ref{eq4.19}) and
(\ref{eq4.17}) vanishes if there is no public disclosure
requirement, i.e., `public disclosure' also makes both the insider
and the market makers trade a larger volume and thus lead to an
increase in total trading volume.

\subsection{Properties of the sequential equilibrium }

\quad\quad In order to analyze the properties of the equilibrium, we
develop an algorithm that analytically solve the model's unique
equilibrium,{\footnote{The algorithm here is different form that of
Holden and Subrahmanyam (1992).}} as described in the following
proposition.
\begin{proposition}\label{pro4.3}
Let $\lambda_n=a_n\frac{\sqrt{\Sigma_{n-1}}}{\sigma_{\mu}}$,
$\alpha_n=b_n\frac{\sigma_{\mu}}{\sqrt{\Sigma_{n-1}}}$,
$\omega_n=c_n\frac{\sigma_{\mu}}{\sqrt{\Sigma_{n-1}}}$, $\forall
n=1,2,\cdots N.$ For $0<K<2,$ the solution of the difference
equation system in Proposition $\ref{pro4.1}$ is given by starting
the boundary condition $b_N=0$,  $c_N=0$,
$a_N=\frac{\sqrt{K(2-K)}}{2}$, and iterating backward for $a_{N-1},
\cdots, a_1$, $b_{N-1}, \cdots, b_1$ and $c_{N-1}, \cdots, c_1$ by
using the following equation, for $n=1,2\cdots, N-1,$
\begin{equation}\label{eq4.22}
b_{n-1}=q_n\left\{1+\frac{K^3q_n}{[(2-K)q_n-z_n](2q_n-z_n)^2}\right\}^\frac{1}{2},
\end{equation}
\begin{equation}\label{eq4.23}
c_{n-1}=z_n\left[1+\frac{K^3q_n}{[(2-K)q_n-z_n][2q_n-z_n)^2}\right]^\frac{1}{2},
\end{equation}
\begin{equation}\label{eq4.23'}
a_n=\frac{K^2b_n}{(2b_n-c_n)^2},
\end{equation}
where
\begin{equation}\begin{aligned}\label{eq4.24}
q_n=&\left[1-\frac{K^4b_n^2}{[(2-K)b_n-c_n](2b_n-c_n)^3}\right]
\frac{K^4b_n^2-(2b_nK-2b_n+c_n)(2b_n-Kb_n-c_n)(2b_n-c_n)^2}{K^2b_n(2b_n-c_n)(2b_n-Kb_n-c_n)}\\
&+b_n \left[1-\frac{K^3b_n }{[(2-K)b_n-c_n](2b_n-c_n)^3}\right]^2,
\end{aligned}
\end{equation}
\begin{equation}\begin{aligned}\label{eq4.25}
z_n=&\frac{K^2b_n(1-K)}{(2b_n-c_n)(2b_n-Kb_n-c_n)}-\left[2-K-\frac{K^4b_n^2}{[(2-K)b_n-c_n](2b_n-c_n)^3}\right]
\frac{(2b_nK-2b_n+c_n)(2b_n-c_n)}{K^2b_n}\\
&+c_n \left[1-\frac{K^3b_n }{[(2-K)b_n-c_n](2b_n-c_n)^2}\right].
\end{aligned}
\end{equation}

Then starting from the exogenous values $\Sigma_0$ and
$\sigma_{\mu}^2$, iterate forward for each of the following
variables in the order listed

\begin{equation}\begin{aligned}\label{eq4.26}
\Sigma_n=\left(1-\frac{Ka_n}{2b_n-Kb_n-c_n}\right)\Sigma_{n-1},
\end{aligned}
\end{equation}
\begin{equation}\begin{aligned}\label{eq4.26,}
\beta_n=\frac{K^4b_n^2-(2b_nK-2b_n+c_n)(2b_n-Kb_n-c_n)(2b_n-c_n)^2}
{2K^2b_n(2b_n-Kb_n-c_n)^2}\frac{\sigma_{\mu}}{\sqrt{\Sigma_{n-1}}},
\end{aligned}
\end{equation}
\begin{equation}\begin{aligned}\label{eq4.27}
\eta_n= \frac{K }{2b_n
-c_n}\frac{\sqrt{\Sigma_{n-1}}}{\sigma_{\mu}},
\end{aligned}
\end{equation}
\begin{equation}\begin{aligned}\label{eq4.28}
\theta_n= \frac{2b_nK-2b_n+c_n }{a_n(2b_n
-c_n)}\frac{\sigma_{\mu}}{\sqrt{\Sigma_{n-1}}},
\end{aligned}
\end{equation}
\begin{equation}\begin{aligned}\label{eq4.29}
\gamma_n= -\frac{2b_nK-2b_n+c_n }{ 2b_n -c_n },
\end{aligned}
\end{equation}
\begin{equation}\begin{aligned}\label{eq4.30}
\gamma_n'= -\frac{2b_nK-2b_n+c_n }{Kb_n },
\end{aligned}
\end{equation}
\begin{equation}\begin{aligned}\label{eq4.31}
\sigma_{z_n}^2= \frac{Kb_n }{2b_n -Kb_n-c_n }\sigma_{\mu}^2-
\frac{K^4b_n^2 }{(2b_n -Kb_n-c_n)^2(2b_n-c_n)^2 }\sigma_{\mu}^2,
\end{aligned}
\end{equation}
for all $n=1,2,\cdots,N,$ and when $n=N$, the boundary condition is
given by Eqs. $(\ref{eq4.13'})$ and $(\ref{eq4.14'})$.
\end{proposition}
\begin{proof} See the Appendix.
\end{proof}

Using  formulas in the proposition we generate a series of numerical
simulations. We first compare the interesting parameters
$\lambda_n$, $\Sigma_n$, $E_1[\Delta\tilde{\pi}_n]$ and
$\sigma_{z_n}^2$ in our model with those of Huddart, Hughes and
Levine (2001) and Wang (1998).
 As in Kyle (1985), the parameters $\Sigma_n$ and
$\lambda_n$ are the inverse measures of price efficiency and the
market depth, respectively.  $\Delta\tilde{\pi}_n$ denotes the
profits of the $n$-th auction, i.e.,
$\Delta\tilde{\pi}_n=\tilde{\pi}_{n+1}-\tilde{\pi}_n.$ And
$\sigma_{z_n}^2$ denotes the variance of the noise that the insider
put when he plays mixed strategy in the $n$-th period.
\begin{figure}
\centering \mbox{ \subfigure[Liquidity parameters, $\lambda_n$, when
the insider must disclosure each trade ex-post, are plotted over
time for different values of $K$, $K=0.5$, $K=0.8$, $K=1.0$,
$K=1.2$,
$K=1.8$.]{\includegraphics[width=.47\textwidth]{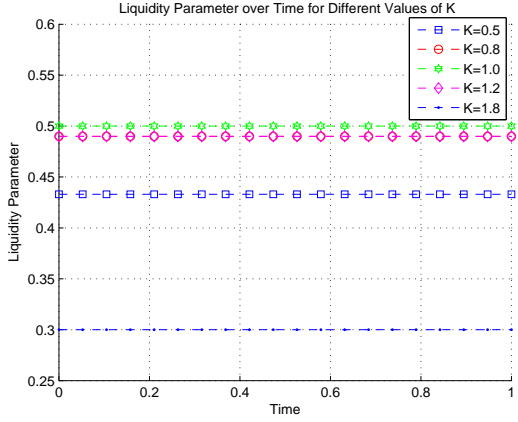}
 }\qquad
 \subfigure[The
liquidity parameter at each auction is plotted for different value
of $K$, $K=0.5$, $K=1.0$, $K=1.2$. $K(without)$ in the figure means
the $K$ in the model that no public disclosure of insider trade is
required, and $K(with)$ means the $K$ in the  model that  the
insider must disclose each trade ex post.]
 {\includegraphics[width=.47\textwidth]{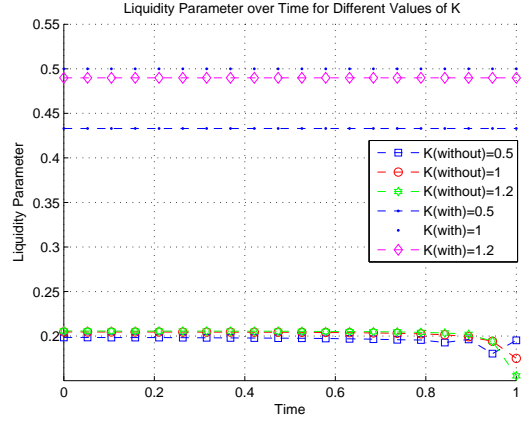}}}
 \caption{\small Numeric solutions of the liquidity parameters $\lambda_{n}$
 with one unit of initial variance of
information.}
\end{figure}
\begin{figure}
\centering \mbox{ \subfigure[The Error variances of price,
$\Sigma_n$, when the insider must disclosure each trade ex post, are
plotted over time for different values of $K$, $K=0.5$, $K=0.8$,
$K=1.0$, $K=1.2$,
$K=1.8$.]{\includegraphics[width=.47\textwidth]{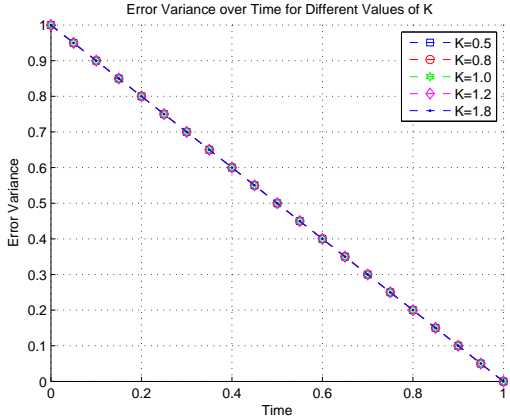}
 }\qquad
 \subfigure[The
error variance of price  at each auction is plotted for different
value of $K$, $K =0.5$, $K=1$, $K=1.2$. K(without) in the figure
means the $K$ in the model that no public disclosure of insider
trade is required, and K(with) means the $K$ in the  model that  the
insider must disclose each trade ex-post.]
 {\includegraphics[width=.47\textwidth]{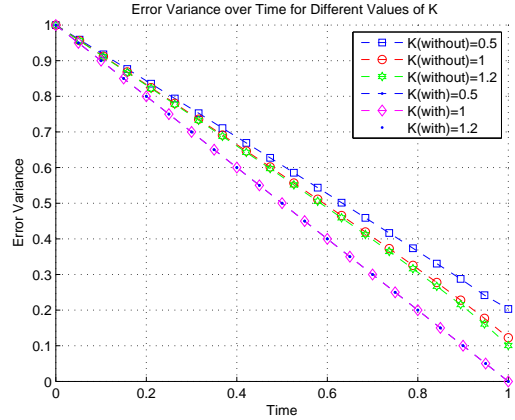}}}
 \caption{\small Numeric solutions of the error variance of price,
 $\Sigma_{n}$,
 with one unit of initial variance of
information.}
\end{figure}

Figure $1$ and Figure $2$ plot the dynamic behavior of the liquidity
parameter $\lambda_n$ and the error variance of price $\Sigma_n$,
respectively, by holding constants $N=20$, $\Sigma_0=1$ and
$\sigma_{\mu}^2=1$ fixed and varying the belief parameter $K=0.5,
0.8, 1.0, 1.2$, or $1.8.$  Among them, the subfigures $1(a)$ and
$2(a)$ plot the parameters $\lambda_n$ and $\Sigma_n$ in our model
respectively for different values $K$, while the subfigures $1(b)$
and $2(b)$ contrast the parameters $\lambda_n$ and $\Sigma_n$,
respectively, (i) when the insider must disclose each trade ex-post
and (ii) when no such disclosures are made.

Figure $1(a)$ shows the trajectory of the market maker's price
adjustment $\lambda_n$ in our model for varying belief parameters.
It indicates that the liquidity parameter $\lambda_n$ is constant
for any values of $K$. This result is consistent with  that of
Huddart, Hughes and Levine (2001). It also indicates that the market
depth, measured by $\frac{1}{\lambda_n}$,  is positively related to
the degree of heterogeneity, measured by $|K-1|$. Figure $1(b)$
indicates that ``public disclosure'' not only leads to constant
market depth for any values of $K$, but also makes the adverse
selection (measured by $\lambda_n$) higher than the case of without
public disclosure requirements. This is because under the
requirements of disclosure the information content of the order flow
is high.

The  pattern of $\lambda_n$ is consistent with the released speed of
the information of the insider, measured by $\Sigma_n$, which is
plotted in Figure $2$.  Figure $2(a)$ indicates that when the
insider must disclose, the value of  $\Sigma_n$ following each
disclosure declines linearly over time independent of periods and
the values of $K$. However, Figure $2(b)$ indicates that less
information asymmetry is present in the market with more aggressive
informed trading, measured by the parameter $K$, without public
disclosure requirements.
\begin{figure}
\centering \mbox{ \subfigure[Expected profit,
$E(\Delta\tilde{\pi}_{n})$, when the insider must disclosure each
trade ex-post, are plotted over time for different values of $K$,
$K=0.5$, $K=0.8$, $K=1.0$, $K=1.2$,
$K=1.8$.]{\includegraphics[width=.47\textwidth]{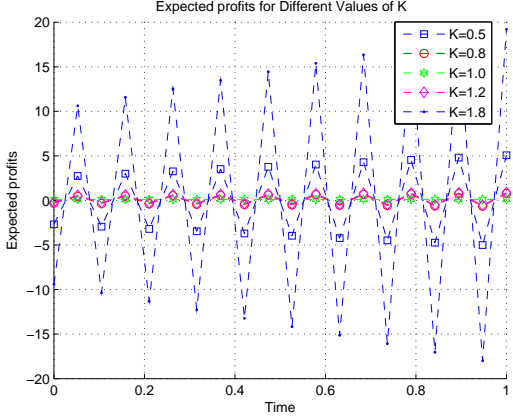}
 }\qquad
 \subfigure[ Expected profit $E(\Delta\tilde{\pi}_{n})$ at each auction is plotted for different value of $K$,
$K =0.5$, $K=1$, $K=1.2$. K(without) in the figure means the $K$ in
the model that no public disclosure of insider trade is required.]
 {\includegraphics[width=.47\textwidth]{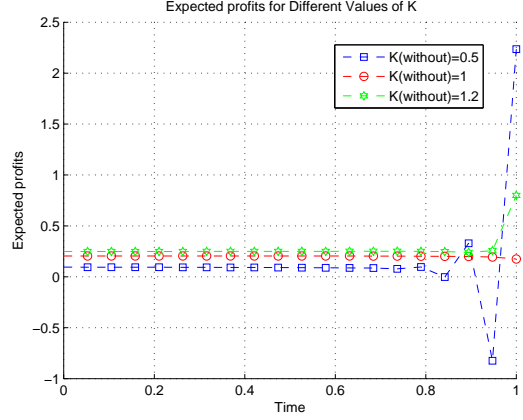}}}
 \caption{\small   Numeric solutions of the expected profit $E(\Delta\tilde{\pi}_{n})$
 with one unit of initial variance of
information.}
\end{figure}

\begin{figure}[htbp]
\begin{minipage}[c]{1.0\textwidth}
\centering
\includegraphics[width=4.5in]{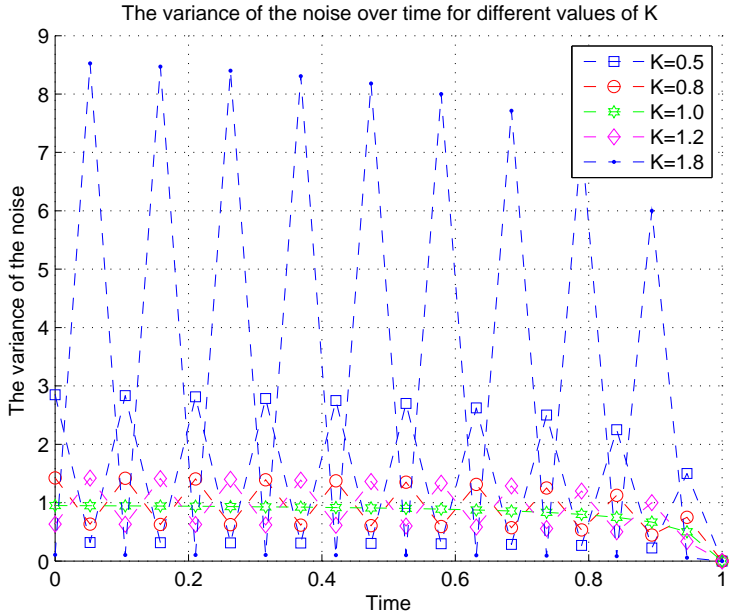}
\caption{\small The variances of the noise that the insider play,
$\sigma_{z_n}^2$, are plotted over time for different values of $K$,
$K=0.5$, $K=1.0$, $K=1.2$, $K=1.5$, $K=1.8$.}
\end{minipage}
\end{figure}

In order to see the differences between the heterogeneous priors
case to the common priors case, and the cases with and without
public disclosure requirements clearly, we plot the dynamic behavior
of $E_1[\Delta\tilde{\pi}_n]$ of the two cases, respectively, in
Figure $3$. Figure $3(a)$ shows that the expected profits of the
rational insider are constant over trading rounds. This is the
result of Huddart, Hughes and Levine (2001). An interesting result
can be get from Figure $3(a)$  is that the insider's profits
fluctuate greatly if the insider is underconfident or overconfident.
Specifically, the sign of the expected profit is alternating to
ensure the last sign of the last period is not negative. Moreover,
the dynamic pattern in Figure $3(a)$ indicates that the fluctuation
of the expected profits is positively related to the degree of
heterogeneity, measured by $|K-1|$. Even though ``disclosure
requirement'' makes the profits of some trading rounds are negative,
the insider trading can also profit from the whole trading process.
While without the public disclosure of insider trades, the expected
profit at each period is almost constant in all early auctions and
becomes significant only in the last few ones.

Figure $4$ indicates that the fluctuation of the noise's variance is
bigger at the beginning few periods and becomes smaller gradually if
the the insider is irrational. The degree of noise variance's
fluctuation
 is positively related to the degree of
heterogeneity, measured by $|K-1|$.  Figure $4$ also  indicates that
more overconfident insider  puts an smaller noise  at the first
period and then put a bigger one, while the more under confident
insider does the opposite.

\begin{figure}
\centering \mbox{ \subfigure[The intensities of trading on private
information, measured by $\beta_n$, are plotted  over time for
different values of $K$, when the insider must disclosure each trade
ex-post.
]{\includegraphics[width=.47\textwidth]{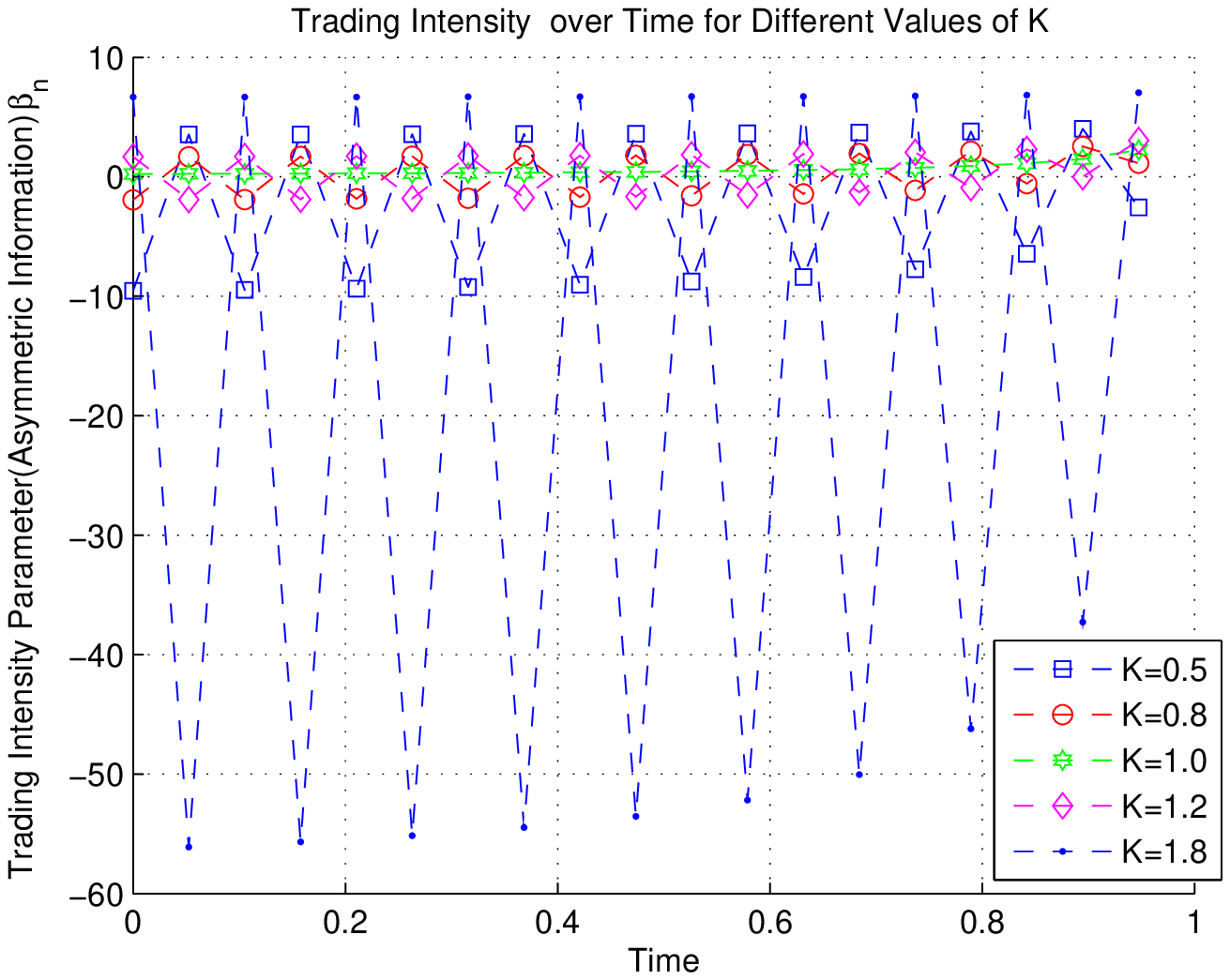}
 }\qquad
 \subfigure[ The intensity of trading on private
information at each auction is plotted for different value of $K$,
$K =0.5$, $K=1$, $K=1.2$. $K(without)$ in the figure means the $K$
in the model that no public disclosure of insider trade is required.
]
 {\includegraphics[width=.47\textwidth]{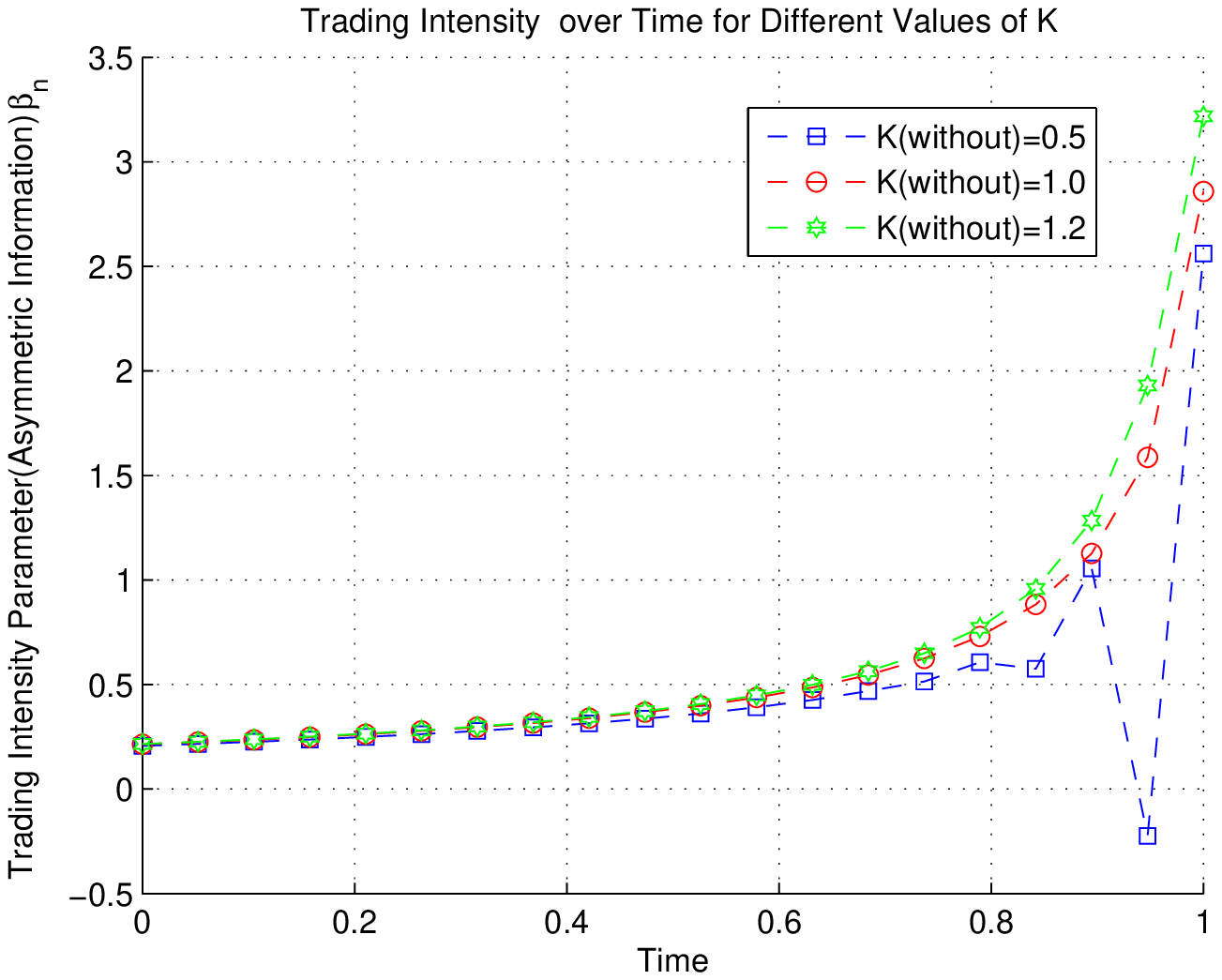}}}
 \caption{\small   The intensity of trading on information, $\beta_n$, over time
for different values of $K$ are plotted.}
\end{figure}

\begin{figure}
\centering \mbox{ \subfigure[The intensities of trading on
heterogeneous prior beliefs, measured by $\theta_n$, are plotted
over time for different values of $K$, when the insider must
disclosure each trade ex-post. The number of auctions is fixed at
$N=20$ and the error variance  at each auction is plotted for
different value of $K$, $K=0.5$, $K=0.8$, $K=1.0$, $K=1.2$,
$K=1.8$.]{\includegraphics[width=.47\textwidth]{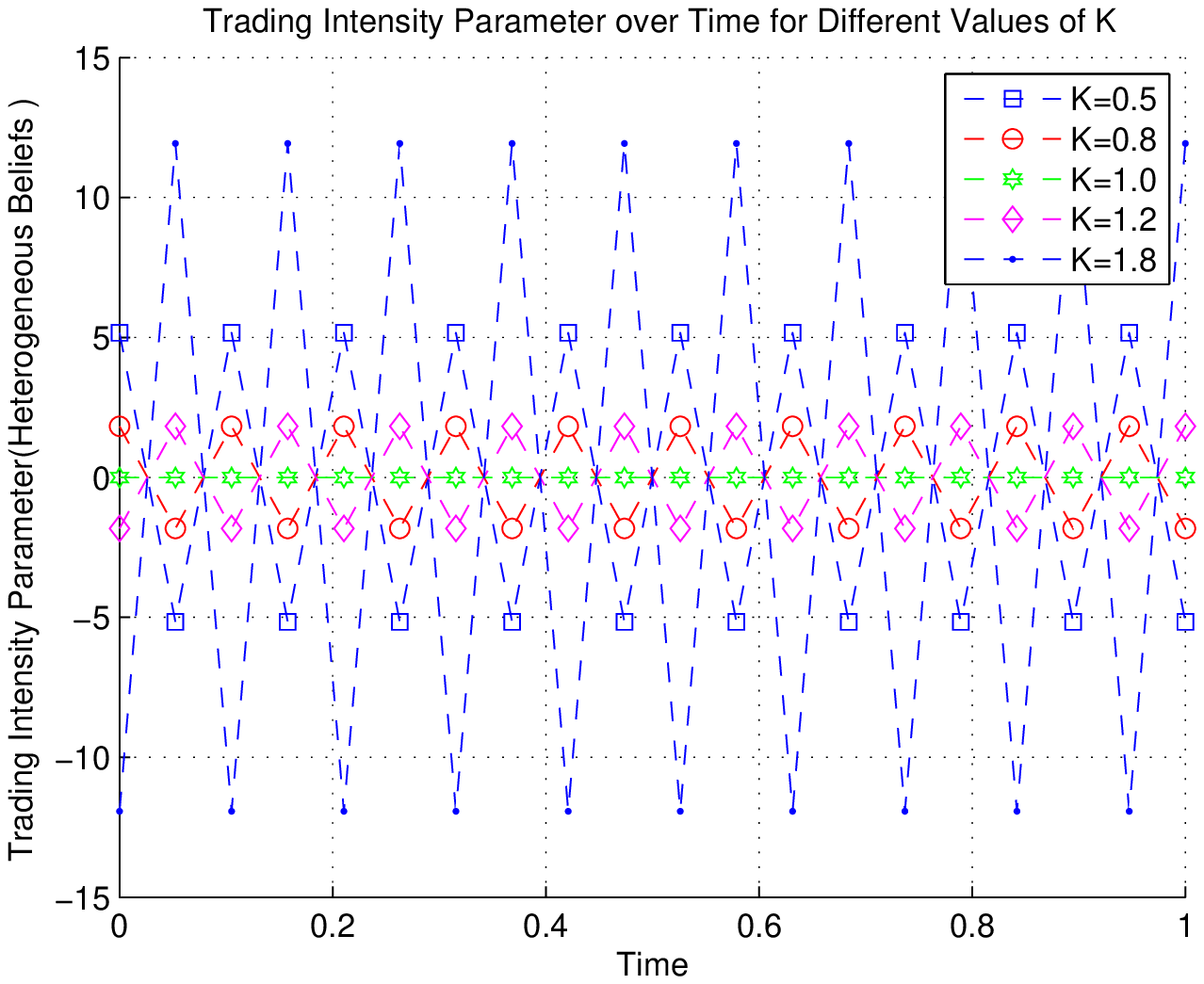}
 }\qquad
 \subfigure[ The intensities of trading on heterogeneous prior beliefs
  at each auction are plotted for different value of $K$,
$K =0.5$, $K=1$, $K=1.2$. K(without) in the figure means the $K$ in
the model that no public disclosure of insider trade is required. ]
 {\includegraphics[width=.47\textwidth]{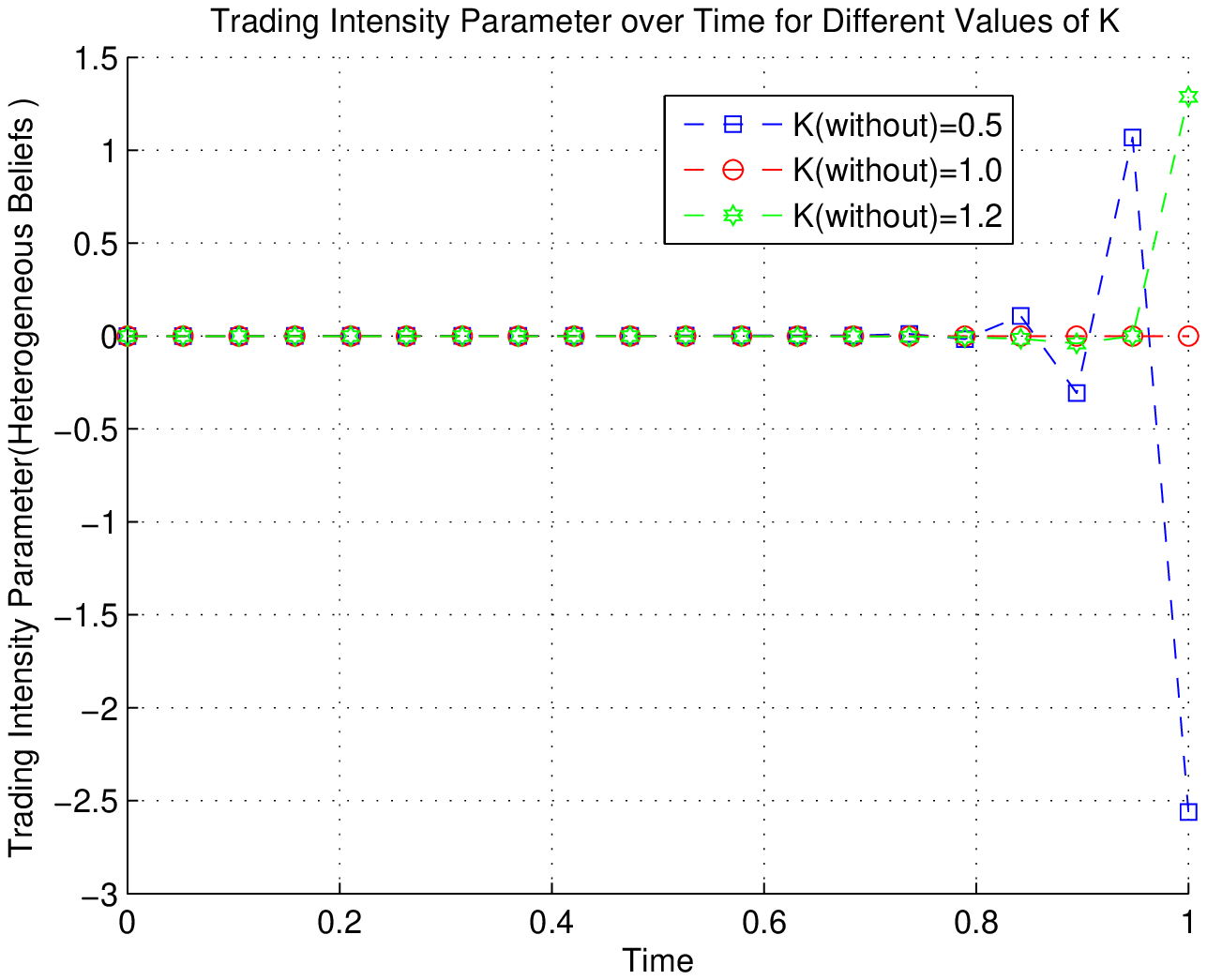}}}
 \caption{\small   The intensity of trading on heterogeneous prior beliefs, $\theta_n$, over time
for different values of $K$ are plotted.}
\end{figure}

\begin{figure}
\centering \mbox{ \subfigure[The heterogeneity parameters,
$\gamma_n$,are  plotted over time for different values of $K$, when
the insider must disclosure each trade ex post. The number of
auctions is fixed at $N=20$ and the error variance  at each auction
is plotted for different value of $K$, $K=0.5$, $K=0.8$, $K=1.0$,
$K=1.2$,
$K=1.8$.]{\includegraphics[width=.47\textwidth]{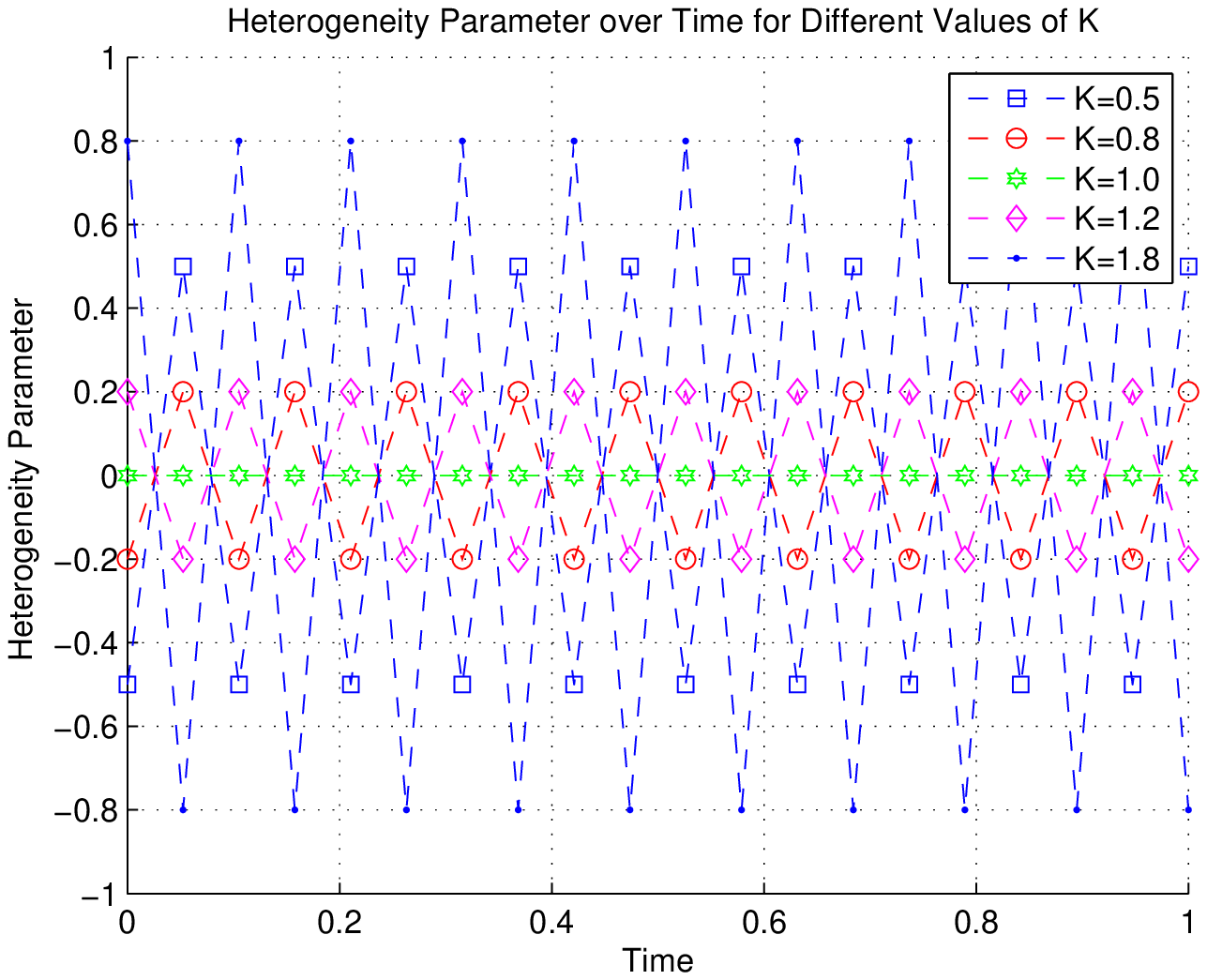}
 }\qquad
 \subfigure[ The heterogeneity parameter,
$\gamma_n$,
  at each auction is plotted for different value of $K$,
$K =0.5$, $K=1$, $K=1.2$. $K(without)$ in the figure means the $K$
in the model that no public disclosure of insider trade is required.
]
 {\includegraphics[width=.47\textwidth]{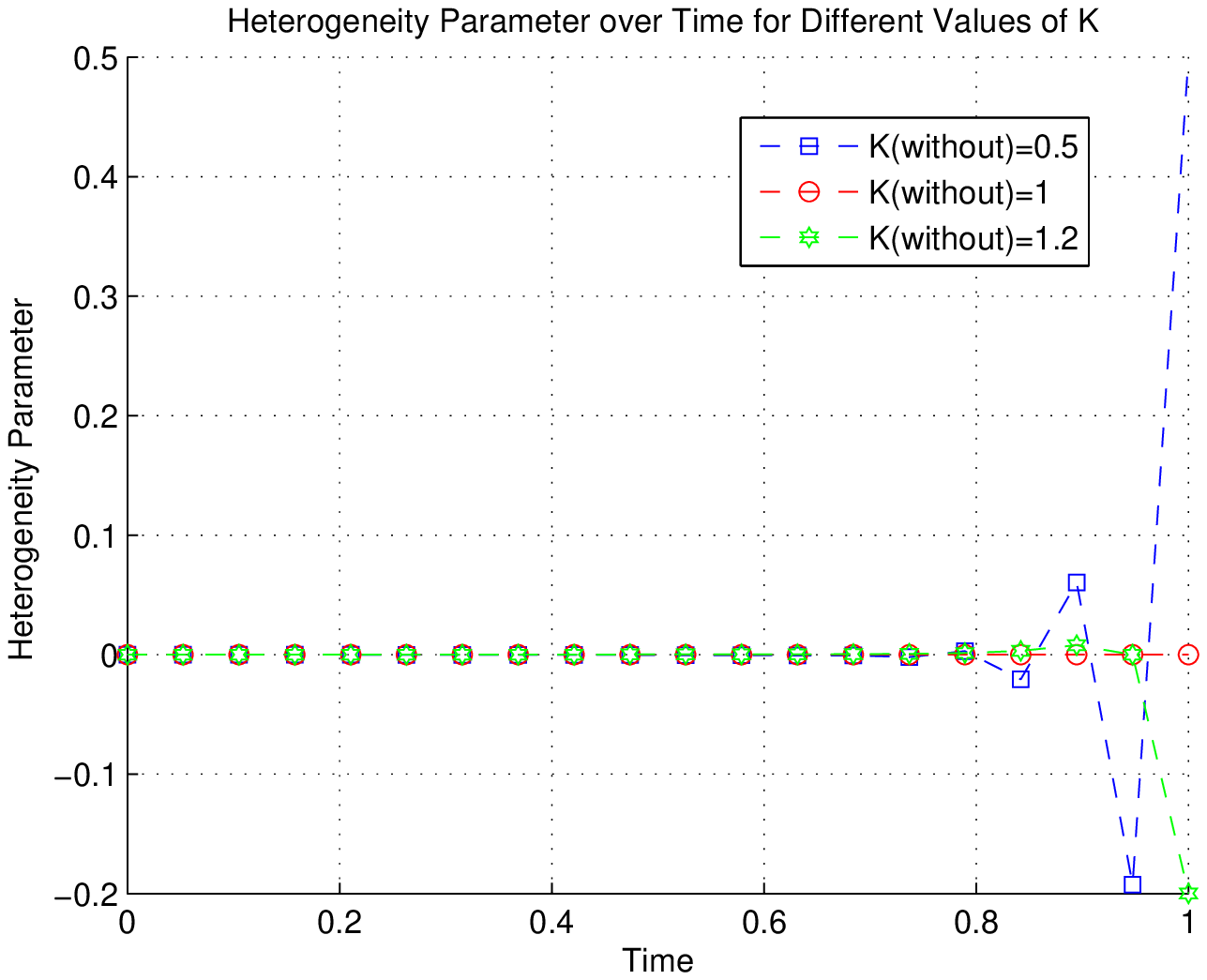}}}
 \caption{\small   The intensity of trading on heterogeneous prior beliefs, $\theta_n$, over time
for different values of $K$ are plotted.}
\end{figure}

\begin{figure}[htbp]
\begin{minipage}[c]{1.0\textwidth}
\centering
\includegraphics[width=4.5in]{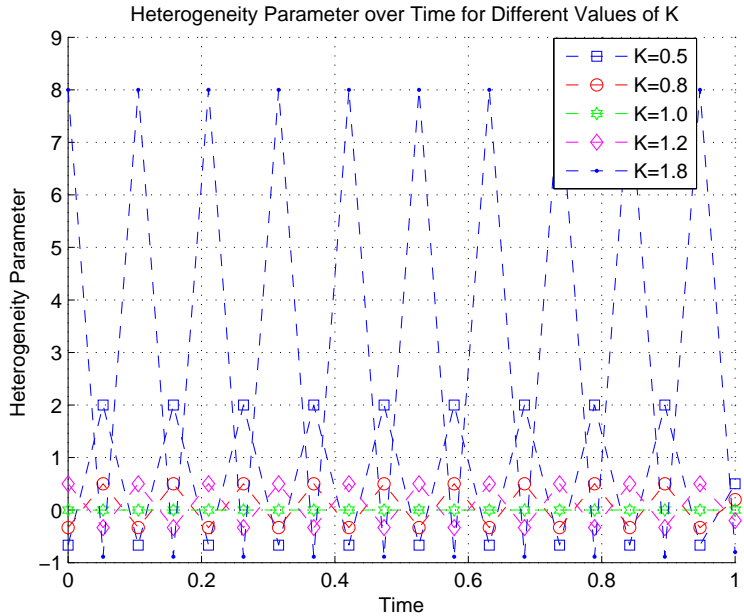}
\caption{ \small The heterogeneous parameters, $\gamma_n'$, over
time for different values of $K$ are plotted, when the insider must
disclosure each trade ex post. The number of auctions is fixed at
$N=20$ and the liquidity parameter at each auction is plotted for
different value of $K$, $K=0.5$, $K=1.0$, $K=1.2$, $K=1.5$,
$K=1.8$.}
\end{minipage}
\end{figure}

\begin{figure}
\centering \mbox{ \subfigure[The expected total trading volume,
$Vol_n$, over time for different values of $K$ are plotted, when the
insider must disclosure each trade ex post. The number of auctions
is fixed at $N=20$ and the liquidity parameter at each auction is
plotted for different value of $K$, $K=0.5$, $K=0.8$, $K=1.0$,
$K=1.2$,
$K=1.8$.]{\includegraphics[width=.47\textwidth]{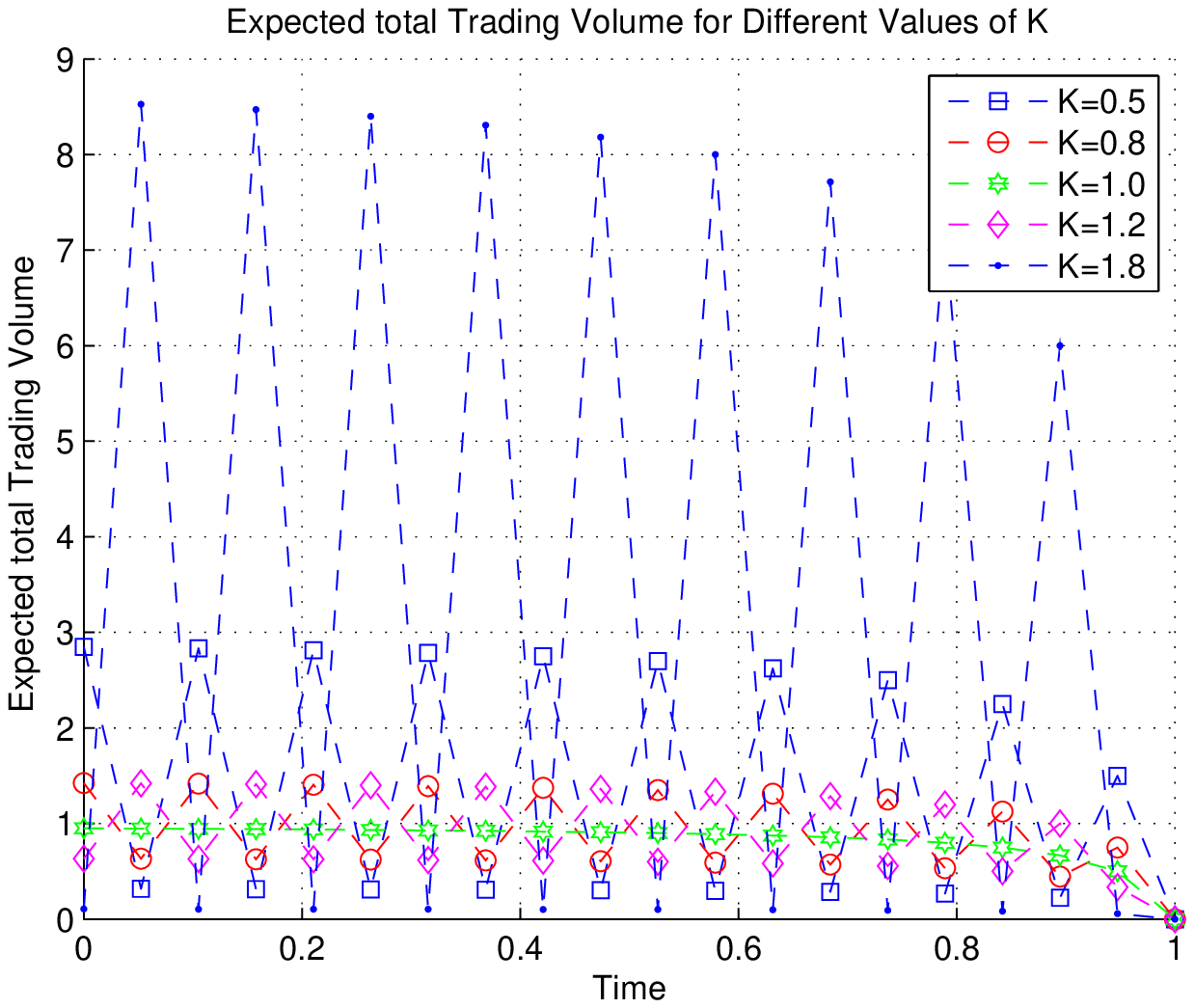}
 }\qquad
 \subfigure[The expected total trading volume at each auction is plotted for different value
of $K$, $K =0.5$, $K=1.0$, $K=1.2$. $K(without)$ in the figure means
the $K$ in the model that no public disclosure of insider trade is
required.]
 {\includegraphics[width=.47\textwidth]{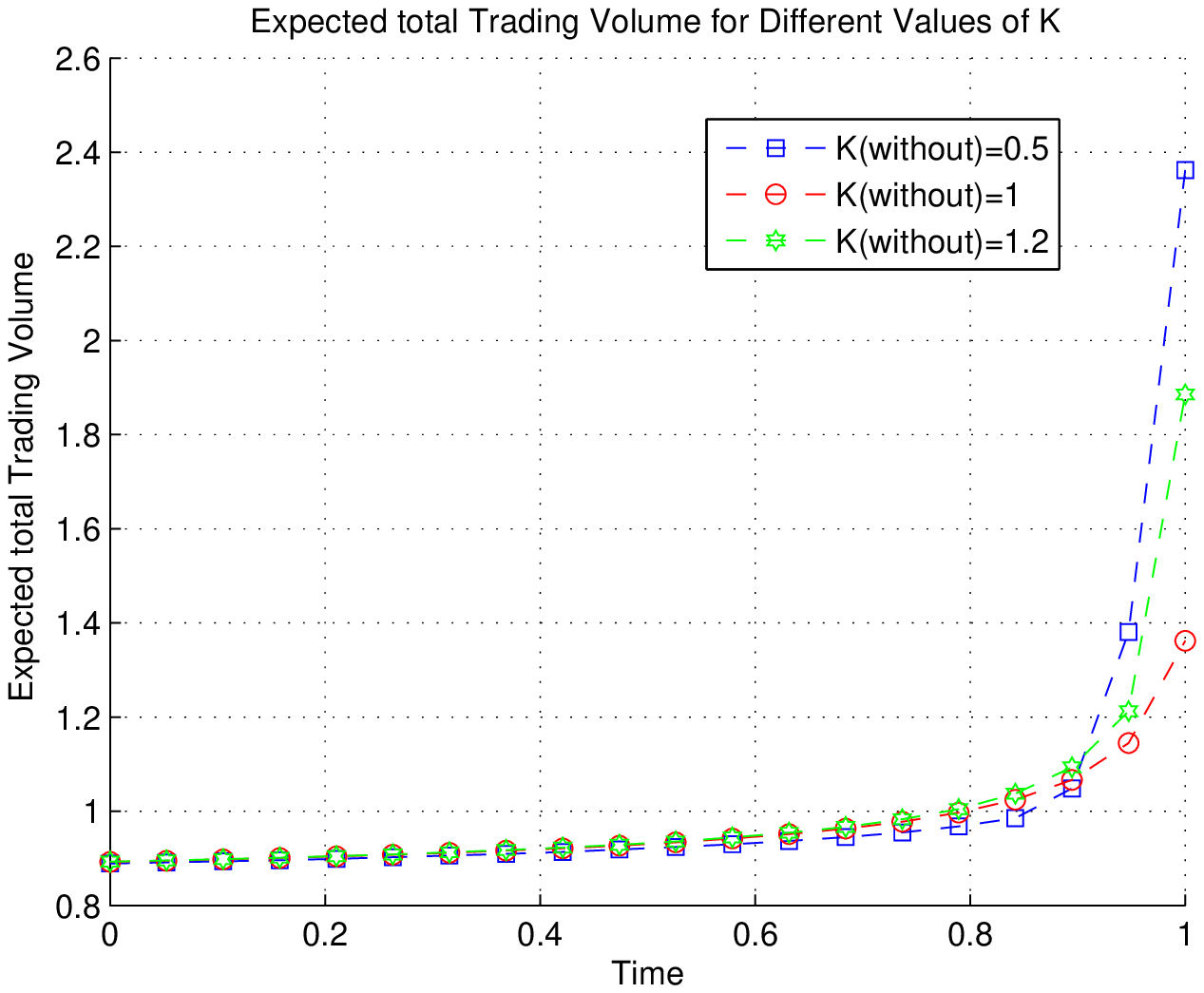}}}
 \caption{\small Numeric solutions of the expected total trading volume $Vol_{n}$
 with one unit of initial variance of
information.}
\end{figure}

Next, we compare parameters $\beta_n$, $\theta_n$, $\gamma_n$
$\gamma_n'$ and $Vol_n$ between the case of with and without
disclosure of insider trades for varying numbers of trading rounds,
by holding constants $N=20$, $\Sigma_0=1$ and $\sigma_{\mu}^2=1$
fixed and varying the belief parameter $K= 0.5$, $0.8$, $1.0$,
$1.2$, or $1.8.$
 As in Wang
(1998), ${\beta_n}$ measures the trading intensity on the asymmetric
information, $\theta_n$ measures the trading intensity on
heterogeneous beliefs, $\gamma_n$ and $\gamma_n'$ are heterogeneity
parameters,  and $Vol_n$ measures the total expected trading volume.

Figure $5(a)$ indicates that under the disclosure requirement the
irrational  informed trader's intensity of trading on asymmetric
information, measured by ${\beta_n}$, fluctuates greatly during all
early auctions and becomes subdued gradually in the last few rounds
of trades.  The fluctuation is also positively related to the degree
of the heterogeneity. This result is intuitive since the more
heterogeneous informed trader is the more he will be influenced by
the ``trade disclosure'' and more aggressively he will trade. Figure
$5(a)$ also indicates if the insider is overconfident the intensity
of trading is positive at the first period and negative at the
second period, and the sigh changes alternatively, while the
underconfident insider does the opposite. This result is very
differently from the case of no disclosure requirement, which is
given in Figure $5(b)$.

Figure $6(a)$  shows  that the irrational  informed trader's
intensity of trading on heterogeneous prior beliefs, measured by
${\theta_n}$, also fluctuates greatly  during all early auctions and
the fluctuation is also positively related to the degree of
heterogeneity, but the direction of fluctuation is opposite to that
of $\beta_n$. That is to say if the irrational  insider puts a
positive weight on asymmetric information this period, then he puts
a negative weight on heterogeneous prior beliefs, and the case is
inverse in the next period.   While Figure $6(b)$ shows that the
informed trader trades on differences in beliefs only in the last
few auctions since the intensity of trading on heterogeneous prior
beliefs is negligible in all early auctions, which is the result of
Wang (1998). Comparing the Figure $5(a)$ with Figure $5(b)$, and
$6(a)$ with Figure $6(a)$ we know that `trade disclosure' has a big
influence on the insider's strategy, whenever the insider is
underconfident or overconfident.

Figure $7$ and Figure $8$ plot the heterogeneity parameters
$\gamma_n$ and $\gamma_n'$, respectively.  They indicate that
$\gamma_n$ and $\gamma_n'$ all also  fluctuate according to the
degree of heterogeneity. This pattern is consistent with the
informed trader's strategies. Market makers can correctly predict
that ``heterogeneous prior beliefs'' and ``trade disclosure'' have
 big impacts on  the trader's strategies at each auction,  market makers
 choose a non-zero $\gamma_n$ to account for the adverse selection
 problem properly,
 and adjust $\gamma_n$ to $\gamma_n'$ correctly after observing the insider's trading. This intuition is further
confirmed by the fact that the patterns in Figure $7(a)$  and Figure
$8$ are  exactly the opposites of the pattern in Figure $6(a)$.

Figure $9(a)$ indicates  that the total expected trading volume
$Vol_n$, is positively related to the degree of heterogeneity,
measured by $|K-1|.$ The fluctuation is  greatly during all early
auctions and becomes subdued gradually in the last few rounds of
trades,  this pattern is consistent with that of the trading
strategy of the  insider. Comparing Figure $9(a)$ with Figure
$9(b)$, we how that  under the requirement of public disclosure, the
expected trading volume is dramatically big. This result vividly
confirms Proposition \ref{pro4.2} which shows that not only
``heterogeneous prior beliefs'' but also ``public disclosure'' lead
to a larger volume.

\section{Conclusions}
\quad\quad In this paper, we have characterized the dynamically
optimal trading strategies of an irrational insider under ex-post
disclosure requirement, and addressed the impacts on the financial
markets.

In contrast to the case of an irrational  insider who restricts his
intensity of trading on private information, during all early
auctions and becomes aggressive only in the last few rounds of
trades in a setting under no disclosure requirement, we show that
the irrational insider under disclosure requirement employs a mixed
strategy and trades more aggressive on private information from the
beginning to the end. In particular, under disclosure requirements,
insider puts the weights on asymmetric information and heterogeneous
prior beliefs are opposite in sign, and the sighs change
alternatively in the next period.   Since on one hand the irrational
insider is overconfident or underconfident about his signal and on
the other hand he wants to dissimulates his information, he trades
more  aggressively than he rational does. The market makers,
realizing the heterogeneity and that a part of total order flow is
due to aggressive behavior, increases market depth. Also, under
disclosure requirement, the market depth is positive related to the
degree of the heterogeneity.

Our model indicates that despite the irrational insider dissembles
his information by adding a random component to his trades, the
information is reflected more rapidly in price with the disclosure
of insider trades than without. An interesting find is that ``public
disclosure'' makes the revelation speed of the information are the
same, whenever the insider is overconfident or underconfident. In
equilibrium, we also show that while ``public disclosure'' may lead
to negative profits at some trading rounds, insider trading remains
profitable from the whole trading time. The irrational insider
trades to make sure his profit at the last period is positive.
Furthermore, the heterogeneity beliefs and public disclosure all
lead to larger trading volume. The co-existence of the heterogeneity
beliefs and public disclosure makes the trading volume fluctuate
greatly, and the fluctuation is positive related to the degree of
the insider's heterogeneity. This result can explain the trading
fluctuation in the real finance in some sense.

\section*{Appendix}

{\bf Proof of Proposition \ref{pro3.1}}: Let the $N$ in Theorem $1$
of Wang (1998) equal $2$ and $\Delta t_n$ equal $1$ $(n=1,2)$, we
have the following relationships:
$$\alpha_2=\omega_2=\phi_2=\delta_2=0,~~\beta_2=\frac{1}{2\lambda_2},
~~\theta_2=-\frac{\gamma_2}{\lambda_2},$$
$$\lambda_2=\frac{[(1+\gamma_2)\beta_2+\theta_2]\Sigma_2}{\sigma_{\mu}^2},~~
\gamma_2=1-K, ~~\Sigma_2=(1+\gamma_2)(1-\lambda_2\beta_2)\Sigma_1,$$
$$\alpha_1=\frac{(1+\gamma_2)^2}{4\lambda_2},~~~\omega_1=\frac{(2-K)\gamma_2}{\lambda_2},~~~
 \phi_1=-\frac{\gamma_2}{\lambda_2},~~~\delta_1=0,$$
$$\beta_1=\frac{1-2\alpha_1\lambda_1}{2\lambda_1(1-\alpha_1\lambda_1)},
~~~\theta_1=-\frac{\gamma_1}{\lambda_1},~~
\lambda_1=\frac{[(1+\gamma_1)\beta_1+\theta_1]\Sigma_1}{\sigma_{\mu}^2},$$
$$\gamma_1=1-K-\omega_1\lambda_1,~~~
\Sigma_1=(1+\gamma_1)(1-\lambda_1\beta_1)\Sigma_0,$$
$$\alpha_0=\frac{(1+\gamma_1)^2}{4\lambda_1(1-\alpha_1\lambda_1)},
~~\omega_0=\omega_1+\frac{(2-K)\gamma_1}{\lambda_1},~~
~~\phi_0=\phi_1-\frac{\gamma_1}{\lambda_1},
~~~\delta_0=\alpha_1\lambda_1^2\sigma_{\mu}^2.$$

By $\gamma_2=1-K$ we can get
$$\beta_2=\frac{1}{2\lambda_2},~~~ \theta_2=\frac{K-1}{\lambda_2},~~~
\lambda_2^2=\frac{K\Sigma_2}{2\sigma_{\mu}^2},~~\Sigma_2=(1-\frac{K}{2})\Sigma_1,$$
$$
\alpha_1=\frac{(2-K)^2}{4\lambda_2},~~\omega_1=\frac{(2-K)(1-K)}{\lambda_2},
 ~~~\phi_1=\frac{K-1}{\lambda_2}.$$
From the second order condition,  $\lambda_2>0$, we can get
$\lambda_2=\sqrt{\frac{K\Sigma_2}{2\sigma_{\mu}^2}}$. From the above
we know that if we get $\Sigma_2$, then we can get $\lambda_2$, and
all the coefficients of the second period is known. But in order to
get $\Sigma_2$, we first solve $\Sigma_1$. Combing the expression of
$\alpha_1$ and $\beta_1$, we get
\begin{equation}\label{eqa1}\beta_1=\frac{[2-m(2-K)^2]}{\lambda_1[4-m(2-K)^2]}.\end{equation}
The expression of $\gamma_1$ and $\omega_1$ imply
\begin{equation}\label{eqa2}\gamma_1=(1-K)[1-m(2-K)],~~
\theta_1=-\frac{\gamma_1}{\lambda_1},~~
\Sigma_1=\frac{2(2-K)[1-m(1-K)]\Sigma_0}{4-m(2-K)^2}.\end{equation}
where $m\doteq\frac{\lambda_1}{\lambda_2}$.

Substitute the expression of $\beta_1,\gamma_1,\theta_1$ and
$\Sigma_1$ into
$\lambda_1=\frac{[(1+\gamma_1)\beta_1+\theta_1]\Sigma_1}{\sigma_{\mu}^2}$,
 we obtain
$$\lambda_1^2=\frac{1}{\sigma_{\mu}^2}\frac{2K(2-K)[2-m(2-K)][1-m(1-K)]\Sigma_0}{[4-m(2-K)^2]^2}.$$
When $4-m(2-K)^2>0$, $[2-m(2-K)][1-m(1-K)]>0$, i.e.
$0<m<\frac{2}{2-K}$
$$\lambda_1=\frac{1}{\sigma_{\mu}}\frac{\sqrt{2K(2-K)[2-m(2-K)][1-m(1-K)]\Sigma_0}}{4-m(2-K)^2}.$$
By the expression of $\Sigma_1$, we get
$$\Sigma_2=\frac{2-K}{2}\Sigma_1=\frac{(2-K)^2[1-m(1-K)]}{4-m(2-K)^2}\Sigma_0,$$
$$\lambda_2=\sqrt{\frac{K\Sigma_2}{2\sigma_{\mu}^2}}=\frac{1}{2\sigma_{\mu}}\sqrt{\frac{2K(2-K)^2[1-m(1-K)]\Sigma_0}{4-m(2-K)^2}}.$$
Since $m\doteq\frac{\lambda_1}{\lambda_2}$ , we have
$$m=2\sqrt{\frac{2-m(2-K)}{(2-K)[4-m(2-K)^2]}}$$
i.e.$$(2-K)^3m^3-4(2-K)m^2-4(2-K)m+8=0,$$ and $m$ satisfy
$0<m<\frac{2}{2-K}$,  $m<\frac{4}{(2-K)^2}.$  From $0<K<2$ we get
$0<m<\frac{2}{2-K}$.  In order to explain the root that satisfy the
condition is unique,  let $f(m)=(2-K)^3m^3-4(2-K)m^2-4(2-K)m+8$. It
is easy to know  $f(0)=8>0$, $f(\frac{2}{2-K})=-\frac{8K}{2-K}<0$,
so the needed root is unique. Q.E.D.\\

{\bf Proof of Proposition \ref{pro3.2}}: In order to get the
proposition, we only need to compute $E_1(\tilde{\pi}_1)$ and
$E_1(\tilde{\pi}_2).$

Using  Eq. (\ref{eq3.4}), we can get
\begin{equation}\label{eq3a}\begin{aligned}E_1(\tilde{\pi}_2)=&E_1\{E_K[\tilde{\pi}_2(\tilde{p}_1^*,\tilde{s})|\tilde{x}_1,\tilde{p_1^*},
\tilde{s}]\}=E_1\left\{\frac{1}{4\lambda_2}[K\tilde{s}-(1+\gamma_2)\tilde{p}_1^*]^2\right\}\\
=&\frac{1}{4\lambda_2}\left\{K^2\Sigma_1+(2K-2)^2E_1[(1+\gamma_1')p_0+\eta_1\tilde{x}_1]^2\right\}\\
=&\frac{K^2\Sigma_1}{4\lambda_2}+\frac{(2K-2)^2}{4\lambda_2}[\eta_1^2(\beta_1+\theta_1)^2\Sigma_0
+(\eta_1\theta_1+\frac{K}{2-K})^2p_0^2+\eta_1^2\sigma_{z_1}^2].
\end{aligned}
\end{equation}
Substitute the expression of $\eta_1, \beta_1, \theta_1$ and
$\sigma_{z_1}^2$ into the above equation, yields
\begin{equation}\label{eq3b}\begin{aligned}&E_1(\tilde{\pi}_2)=
\frac{K\sqrt{K\Sigma_0}\sigma_{\mu}}{2\sqrt{2(2-K)}}+\frac{(2K-2)^2\sigma_{\mu}}{\sqrt{2K(2-K)\Sigma_0}}\times\\
&
\left\{\frac{K\Sigma_0}{2(2-K)\sigma_{\mu}^2}\frac{\sigma_{\mu}^2}{\Sigma_0}\frac{2-K}{2K}\Sigma_0+\left[
\frac{\sqrt{K\Sigma_0}}{\sigma_{\mu}\sqrt{2(2-K)}}\frac{2\sqrt{2}(1-K)\sigma_{\mu}}{\sqrt{K(2-K)\Sigma_0}}+\frac{K}{2-K}\right]^2p_0^2
+\frac{K\Sigma_0}{2(2-K)\sigma_{\mu}^2}\frac{2-K}{2K}\sigma_{\mu}^2\right\}\\
=&\frac{5K^2-8K+4}{2\sqrt{2K(2-K)}}\sigma_{\mu}\sqrt{\Sigma_0}+\frac{(2K-2)^2}{\sqrt{2K(2-K)\Sigma_0}}\sigma_{\mu}p_0^2.
\end{aligned}
\end{equation}
Eq. (\ref{eq3.6})  implies
\begin{equation}\label{eq3c}\begin{aligned}
E_1(\tilde{\pi}_1)=&
E_1\{E_K[\tilde{x}(\tilde{v}-\tilde{p}_1)+\tilde{\pi}_2(\tilde{p}_1^*,\tilde{s})|\tilde{s}]\}\\
=&E_1\{[\beta_1(\tilde{s}-p_0)+\theta_1\tilde{s}+\tilde{z}_1][K\tilde{s}-p_0-\gamma_1p_0-\lambda_1\beta_1(\tilde{s}-p_0)-
\lambda_1\theta_1\tilde{s}-\lambda_1\tilde{z}_1]\}+E_1(\tilde{\pi}_2),
\end{aligned}
\end{equation}
while
\begin{equation}\label{eq3d}\begin{aligned}
&E_1\{[\beta_1(\tilde{s}-p_0)+\theta_1\tilde{s}+\tilde{z}_1][K\tilde{s}-p_0-\gamma_1p_0-\lambda_1\beta_1(\tilde{s}-p_0)-
\lambda_1\theta_1\tilde{s}-\lambda_1\tilde{z}_1]\}\\
=&(K-\lambda_1\beta_1-\lambda_1\theta_1)(\beta_1+\theta_1)(\Sigma_0+p_0^2)+(\beta_1+\theta_1)(\lambda_1\beta_1-1-\gamma_1)p_0^2-
\beta_1(K-\lambda_1\beta_1-\lambda_1\theta_1)p_0^2\\
&-\beta_1(\lambda_1\beta_1-1-\gamma_1)p_0^2-\lambda_1\sigma_{z_1}^2\\
=&\frac{3K-2}{2}\sqrt{\frac{2-K}{2K}}\sqrt{\Sigma_0}\sigma_{\mu}+(K-1)\sqrt{\frac{2K}{2-K}}\frac{\sigma_{\mu}}{\sqrt{\Sigma_0}}p_0^2.
\end{aligned}
\end{equation}
 Eqs. (\ref{eq3c}) and (\ref{eq3d}) imply
\begin{equation}\label{eq3e}\begin{aligned}
&E_1(\tilde{\pi}_1)=
\frac{2K^2}{2\sqrt{2K(2-K)}}\sqrt{\Sigma_0}\sigma_{\mu}+\frac{6K^2-10K+4}{\sqrt{2K(2-K)}}\frac{\sigma_{\mu}}{\sqrt{\Sigma_0}}p_0^2.
\end{aligned}
\end{equation}Q.E.D.

{\bf Proof of Proposition \ref{pro3.3}}: From Proposition
\ref{pro3.2}, it is easy to get
$$\beta_1(1+\gamma_1)+\theta_1=\frac{\sigma_{\mu}}
{\sqrt{\Sigma_0}}\sqrt{\frac{2-K}{2K}},~~\beta_2(1+\gamma_2)+\theta_2=\frac{K}{2\lambda_2}=\frac{\sqrt{2K}\sigma_{\mu}}{\sqrt{(2-K)\Sigma_0}},$$
$$\lambda_1=\lambda_2=\frac{\sqrt{K(2-K)\Sigma_0}}{2\sqrt{2}\sigma_{\mu}},~~
 \Sigma_1=\frac{\Sigma_0}{2}, ~~
\sigma_{z_1}^2=\frac{2-K}{2K}\sigma_{\mu}^2,$$
 and the total expected profit is
$$E_1(\tilde{\pi}_1)=
\frac{2K^2}{2\sqrt{2K(2-K)}}\sqrt{\Sigma_0}\sigma_{\mu}+\frac{6K^2-10K+4}{\sqrt{2K(2-K)}}\frac{\sigma_{\mu}}{\sqrt{\Sigma_0}}p_0^2.
$$
By Proposition $2$ in Huddart, et al.(2001), we get
$$\bar{\beta}_1=
\frac{\sigma_{\mu}}{\sqrt{2\Sigma_0}},~~\bar{\beta}_2=\sigma_{\mu}\sqrt{\frac{2}{\Sigma_0}},~~\bar{\lambda}_1=\bar{\lambda}_2=\frac{1}{2\sigma_{\mu}}\sqrt{\frac{\Sigma_0}{2}},$$
$$\bar{\Sigma}_1=\frac{\Sigma_0}{2}, ~~\bar{\sigma}_{z_1}^2=\frac{1}{2}\sigma_{\mu}^2,$$
 and the total expected profit is
 $\frac{\sigma_{\mu}\sqrt{\Sigma_0}}{\sqrt{2}}.$

 Comparing the corresponding parameters respectively, it is easy to
 get the conclusion.Q.E.D.\\

{\bf Proof of Proposition \ref{pro3.4}}: (i)  By taking $K=0.5$ in
Proposition \ref{pro3.1}, we can have $m=0.9235,$ and
$$\hat{\lambda}_1\approx0.3665\frac{\sqrt{\Sigma_0}}{\sigma_{\mu}},~
~\hat{\lambda}_2\approx0.3969\frac{\sqrt{\Sigma_0}}{\sigma_{\mu}},~
~\hat{\beta}_1\approx-0.1105\frac{\sigma_{\mu}}{\sqrt{\Sigma_0}},~~
\hat{\beta}_2\approx1.2598\frac{\sigma_{\mu}}{\sqrt{\Sigma_0}},~~\hat{\gamma}_1\approx-0.1926,$$
$$\hat{\gamma}_2=\frac{1}{2},~~\hat{\theta}_1\approx0.5255\frac{\sigma_{\mu}}{\sqrt{\Sigma_0}},~~
\hat{\theta}_2\approx-1.2598\frac{\sigma_{\mu}}{\sqrt{\Sigma_0}},~~
\hat{\Sigma}_1\approx0.8401\Sigma_0,~
~\hat{\Sigma}_2\approx0.6301\Sigma_0,$$
$$E(\hat{\pi}_1)\approx0.3814\sigma_{\mu}\sqrt{\Sigma_0}+0.3670\frac{\sigma_{\mu}}{\sqrt{\Sigma_0}}p_0^2,$$
$$E(\hat{\pi}_2)\approx0.2329\sigma_{\mu}\sqrt{\Sigma_0}+0.6298\frac{\sigma_{\mu}}{\sqrt{\Sigma_0}}p_0^2.$$

And taking $K=0.5$ in Proposition \ref{pro3.2}, we get
$$\lambda_1=\lambda_2\approx
0.3062\frac{\sqrt{\Sigma_0}}{\sigma_{\mu}},~~
\eta_1\approx0.4082\frac{\sqrt{\Sigma_0}}{\sigma_{\mu}},~~
\gamma_1=-\gamma_2=-0.5,~~
\beta_1\approx-0.8165\frac{\sigma_{\mu}}{\sqrt{\Sigma_0}},~~\beta_2\approx1.6330\frac{\sigma_{\mu}}{\sqrt{\Sigma_0}}
,$$
$$\theta_1=-\theta_2\approx1.6330\frac{\sigma_{\mu}}{\sqrt{\Sigma_0}}, ~~\Sigma_1=\frac{1}{2}\Sigma_0,$$
$$E(\tilde{{\pi}}_1)\approx0.2041\sigma_{\mu}\sqrt{\Sigma_0}+0.4082\frac{\sigma_{\mu}}{\sqrt{\Sigma_0}}p_0^2,$$
$$E(\tilde{{\pi}}_2)\approx0.5103\sigma_{\mu}\sqrt{\Sigma_0}+0.8165\frac{\sigma_{\mu}}{\sqrt{\Sigma_0}}p_0^2.$$

(ii)  By taking $K=1.5$ in Proposition  \ref{pro3.1}, we can have
$m=1.6257$ and
$$ \hat{\lambda}_1\approx0.5000\frac{\sqrt{\Sigma_0}}{\sigma_{\mu}},~~
 \hat{\lambda}_2\approx0.3076\frac{\sqrt{\Sigma_0}}{\sigma_{\mu}},~~
~\hat{\beta}_1\approx0.8869\frac{\sigma_{\mu}}{\sqrt{\Sigma_0}},~~
\hat{\beta}_2\approx1.6255\frac{\sigma_{\mu}}{\sqrt{\Sigma_0}},~~\hat{\gamma}_1\approx-0.0936,$$
$$\hat{\gamma}_2=-\frac{1}{2},~~
\hat{\theta}_1\approx0.1872\frac{\sigma_{\mu}}{\sqrt{\Sigma_0}},~~
\hat{\theta}_2\approx1.6255\frac{\sigma_{\mu}}{\sqrt{\Sigma_0}},~~
~\hat{\Sigma}_1\approx0.5045\Sigma_0,~
~\hat{\Sigma}_2\approx0.1261\Sigma_0,$$
$$E(\hat{\pi}_1)\approx2.3207\sigma_{\mu}\sqrt{\Sigma_0}+0.9064\frac{\sigma_{\mu}}{\sqrt{\Sigma_0}}p_0^2,$$
$$E(\hat{\pi}_2)\approx1.3253\sigma_{\mu}\sqrt{\Sigma_0}+0.8128\frac{\sigma_{\mu}}{\sqrt{\Sigma_0}}p_0^2$$
And taking $K=1.5$ in Proposition \ref{pro3.2}, we get
$$\lambda_1=\lambda_2\approx
0.3062\frac{\sqrt{\Sigma_0}}{\sigma_{\mu}},~~
\eta_1\approx1.2247\frac{\sqrt{\Sigma_0}}{\sigma_{\mu}},~~
\gamma_1=-\gamma_2=0.5,~~
\beta_1\approx1.3608\frac{\sigma_{\mu}}{\sqrt{\Sigma_0}},~~\beta_2\approx1.6330\frac{\sigma_{\mu}}{\sqrt{\Sigma_0}}
,$$
$$\theta_1=-\theta_2\approx-1.6330\frac{\sigma_{\mu}}{\sqrt{\Sigma_0}}, ~~\Sigma_1=\frac{1}{2}\Sigma_0,$$
$$E(\tilde{{\pi}}_1)\approx1.8371\sigma_{\mu}\sqrt{\Sigma_0}+2.0412\frac{\sigma_{\mu}}{\sqrt{\Sigma_0}}p_0^2,$$
$$E(\tilde{{\pi}}_2)\approx1.3268\sigma_{\mu}\sqrt{\Sigma_0}+0.8165\frac{\sigma_{\mu}}{\sqrt{\Sigma_0}}p_0^2.$$

Comparing the corresponding parameters respectively, it is easy to
 get the conclusion.Q.E.D.

 {\bf Proof of Proposition \ref{pro4.1}}:  For
$N(N\geq2)$, suppose that there exist constants
$\beta_n,\theta_n,\gamma_n,\lambda_n,\gamma_n',\eta_n$ such that
$$\tilde{x}_n=\beta_n(1+\gamma_n)(\tilde{s}-p_{n-1}^*)+\theta_n\tilde{s}+\tilde{z}_n,$$
$$\tilde{p}_n-\tilde{p}_{n-1}^*=\gamma_n\tilde{p}_{n-1}^*+\lambda_n(\tilde{x}_n+\tilde{\mu}_n),$$
$$\tilde{p}_n^*-\tilde{p}_{n-1}^*=\gamma_n'\tilde{p}_{n-1}^*+\eta_n\tilde{x}_n.$$

We will use the backward induction method to prove that the unique
linear equilibrium exists. Suppose that there exist constants
$\alpha_n,\omega_n,\phi_n$ and $\delta_n$  such that
$$\begin{aligned}&E_K[\tilde{\pi}_{n+1}|\tilde{p}_1^*=p_1^*,\cdots,\tilde{p}_{n}^*=p_{n}^*,
\tilde{p}_1=p_1,\cdots,\tilde{p}_{n}=p_n,\tilde{s}=s]\\&=\alpha_{n}(s-p_{n}^*)^2
+\omega_{n}sp_{n}^*+\phi_{n}s^2+\delta_{n}.\end{aligned}$$

Applying the principal of backward induction, we can write
 the insider's last period ($N$th period) optimization problem
 for given $\tilde{x}_{1}=x_{1},\cdots,\tilde{x}_{N-1}=x_{N-1},\tilde{p}_1=p_1,\cdots,
 \tilde{p}_{n-1}=p_{n-1},\tilde{p}_1^*=p_1^*,\cdots,\tilde{p}_{N-1}^*=p_{N-1}^*,\tilde{s}=s$ as
$$\begin{aligned}x_{N} \in
  arg&\max_{x}E_K[\tilde{x}(\tilde{v}-\tilde{p}_N)|\tilde{x}_{1}=x_{1},\cdots,\tilde{x}_{N-1}=x_{N-1},\tilde{p}_1=p_1,\cdots,
 \tilde{p}_{n-1}=p_{n-1},\\&\tilde{p}_1^*=p_1^*,\cdots,\tilde{p}_{N-1}^*=p_{N-1}^*,\tilde{s}=s],
 \end{aligned}
 $$
  where
$$\begin{aligned}
&E_K[\tilde{x}(\tilde{v}-\tilde{p}_N)|\tilde{x}_{1}=x_{1},\cdots,\tilde{x}_{N-1}=x_{N-1},\tilde{p}_1=p_1,\cdots,
 \tilde{p}_{n-1}=p_{n-1},\tilde{p}_1^*=p_1^*,\cdots,\tilde{p}_{N-1}^*=p_{N-1}^*,\tilde{s}=s]\\
 =&E_K[\tilde{x}(\tilde{v}-\tilde{p}_{N-1}^*-\gamma_N\tilde{p}_{N-1}^*-\lambda_N(\tilde{x}+\tilde{\mu}_N))|
 \tilde{x}_{1}=x_{1},\cdots,\tilde{x}_{N-1}=x_{N-1},\tilde{p}_1=p_1,\cdots,
 \tilde{p}_{n-1}=p_{n-1},\\
 &\tilde{p}_1^*=p_1^*,\cdots,\tilde{p}_{N-1}^*=p_{N-1}^*,\tilde{s}=s]\\
=&x[Ks-p_{N-1}^*-\gamma_Np_{N-1}^*-\lambda_Nx].
\end{aligned}
$$
The first order condition implies
\begin{equation}\label{eqa4'}x_N=\frac{1+\gamma_N}{2\lambda_N}(s-p_{N-1}^*)+\frac{1}{2\lambda_N}(K-1-\gamma_N)s,
\end{equation}
and the second order condition is $\lambda_N>0.$

From Eq. (\ref{eqa4'}), we know that
\begin{equation}\label{eqa4''}\beta_N=\frac{1}{2\lambda_N},~~~~~~~\theta_N=\frac{1}{2\lambda_N}(K-1-\gamma_N),
\end{equation}
and
\begin{equation}\label{eqa4'''}\begin{aligned}
&E_K[\tilde{x_N}(\tilde{v}-\tilde{p}_N)|\tilde{x}_{1}=x_{1},\cdots,\tilde{x}_{N-1}=x_{N-1},\tilde{p}_1=p_1,\cdots,
 \tilde{p}_{n-1}=p_{n-1},\\&\tilde{p}_1^*=p_1^*,\cdots,\tilde{p}_{N-1}^*=p_{N-1}^*,\tilde{s}=s]\\
=&\left[\frac{1+\gamma_N}{2\lambda_N}(s-p_{N-1}^*)+\frac{1}{2\lambda_N}(K-1-\gamma_N)s\right]\times\\
&\left[Ks-p_{N-1}^*-\gamma_Np_{N-1}^*-
\frac{1+\gamma_N}{2}(s-p_{N-1}^*)-\frac{1}{2}(K-1-\gamma_N)s\right]\\
=&\frac{1}{4\lambda_N}\left[(1+\gamma_N)(s-p_{N-1}^*)+(K-1-\gamma_N)s\right]^2&\\
=&\frac{1}{4\lambda_N}\left[(1+\gamma_N)^2(s-p_{N-1}^*)^2-2(1+\gamma_N)(K-1-\gamma_N)sp_{N-1}^*+(K+1+\gamma_N)(K-1-\gamma_N)s^2\right],
\end{aligned}
\end{equation}
so
\begin{equation}\label{eqa4a}\alpha_{N-1}=\frac{(1+\gamma_N)^2}{4\lambda_N},~~~~\omega_{N-1}=\frac{-2(1+\gamma_N)(K-1-\gamma_N)}{4\lambda_N},~~~\phi_{N-1}
=\frac{(K+1+\gamma_N)(K-1-\gamma_N)}{4\lambda_N}.
\end{equation}
Since $\gamma_N=-\lambda_N\theta_N$, $\gamma_N'=-\eta_N\theta_N$, by
the market efficient condition, we have
\begin{equation}\label{eqa4b}\begin{aligned}\tilde{p}_N-\tilde{p}_{N-1}^*
=&E_1(\tilde{v}-\tilde{p}_{N-1}^*|
\tilde{p}_1^*=p_1^*,\cdots,\tilde{p}_{N-1}^*=p_{N-1}^*,
\tilde{p}_1=p_1,\cdots,\tilde{p}_{N-1}=p_{N-1},\\&\tilde{x}_1=x_1,\cdots,\tilde{x}_{N-1}=x_{N-1},\tilde{y}_N=y_N)\\
=&
E_1\left[\tilde{v}-\tilde{p}_{n-1}^*|\tilde{x}_N+\tilde{\mu}_N+\frac{\gamma_N}{\lambda_N}p_{N-1}^*\right]\\
=&E_1\left[\tilde{v}-\tilde{p}_{N-1}^*|\frac{\gamma_N}{\lambda_N}\tilde{p}_{N-1}^*+\beta_N(1+\gamma_N)
(\tilde{s}-p_{N-1}^*)+\theta_N\tilde{s}+\tilde{\mu}_N\right]\\
=&E_1\left[\tilde{v}-\tilde{p}_{N-1}^*|(\beta_N(1+\gamma_N)+\theta_N)
(\tilde{s}-\tilde{p}_{N-1}^*)+\tilde{\mu}_N\right],
\end{aligned}
\end{equation}
so
 $$\lambda_N=\frac{(\beta_N(1+\gamma_N)+\theta_N)\Sigma_{N-1}}{(\beta_N(1+\gamma_N)+\theta_N)^2\Sigma_{N-1}+\sigma_{\mu}^2}.$$
Substitute Eq. (\ref{eqa4''}) into the above expression, we have
\begin{equation}\label{eqa4c}\lambda_N=\frac{\frac{K}{2\lambda_N}\Sigma_{N-1}}{\frac{K^2}{4\lambda_N^2}\Sigma_{N-1}+\sigma_{\mu}^2},
\end{equation}
and combing the above with the second order condition, we get
\begin{equation}\label{eqa4d}\lambda_N=\sqrt{\frac{K(2-K)\Sigma_{N-1}}{4\sigma_{\mu}^2}}.
\end{equation}
So the $K$ should satisfied $0<K<2.$  Eq. (\ref{eqa4''}) and
$\gamma_N=-\lambda_N\theta_N$ imply
$$\theta_N=\frac{K-1}{\lambda_N},~~~~~~\gamma_N=1-K.$$
Substitute the above to Eq. (\ref{eqa4a}), we have
\begin{equation}\label{eqa4d'}\alpha_{N-1}=\frac{(1+\gamma_N)^2}{4\lambda_N}=\frac{(2-K)^2}{4\lambda_N},
\end{equation}
\begin{equation}\label{eqa4d''}\omega_{N-1}=\frac{-2(1+\gamma_N)(K-1-\gamma_N)}{4\lambda_N}=\frac{(2-K)(1-K)}{\lambda_N},
\end{equation}
\begin{equation}\label{eqa4d'''}
\phi_{N-1}
=\frac{(K+1+\gamma_N)(K-1-\gamma_N)}{4\lambda_N}=\frac{K-1}{\lambda_N}.
\end{equation}
 Since
\begin{equation}\label{eqa4e}\begin{aligned}\Sigma_{N}=&Var_1(\tilde{v}|
\tilde{p}_1^*=p_1^*,\cdots,\tilde{p}_{N-1}^*=p_{N-1}^*,
\tilde{p}_1=p_1,\cdots,\tilde{p}_{N-1}=p_{N-1},
\\&\tilde{x}_1=x_1,\cdots,\tilde{x}_{N-1}=x_{N-1},\tilde{x}_N=x_N)\\
=&
Var_1\left[\tilde{v}-p_{N-1}^*|\tilde{x}_N+\frac{\gamma_N}{\lambda_N}p_{N-1}^*\right]\\
=&Var_1\left[\tilde{v}-p_{N-1}^*|\frac{\gamma_N}{\lambda_N}p_{N-1}^*+\beta_N(1+\gamma_N)
(\tilde{s}-p_{N-1}^*)+\theta_N\tilde{s}\right]\\
=&Var_1\left[\tilde{v}-\tilde{p}_{N-1}^*|(\beta_N(1+\gamma_N)+\theta_N)
(\tilde{s}-\tilde{p}_{N-1}^*)\right],
\end{aligned}
\end{equation}
we have
$$\Sigma_N=\Sigma_{N-1}-\frac{(\beta_N(1+\gamma_N)+\theta_N)^2\Sigma_{N-1}^2}{(\beta_N(1+\gamma_N)+\theta_N)^2\Sigma_{N-1}}=0.$$
  Since
$\tilde{\pi}_n=\tilde{\pi}_{n+1}+\tilde{x}_n(\tilde{v}-\tilde{p}_n)$£¬
we have
\begin{equation}\label{eqa4}\begin{aligned}
&E_K[\tilde{\pi}_n|\tilde{p}_1^*=p_1^*,\cdots,\tilde{p}_{n-1}^*=p_{n-1}^*,
\tilde{p}_1=p_1,\cdots,\tilde{p}_{n-1}=p_{n-1},\tilde{s}=s]\\
=&\max_xE_K[\tilde{x}(\tilde{v}-\tilde{p}_n)||\tilde{p}_1^*=p_1^*,\cdots,\tilde{p}_{n-1}^*=p_{n-1}^*,
\tilde{p}_1=p_1,\cdots,\tilde{p}_{n-1}=p_{n-1},\tilde{s}=s]\\&+E_K[\alpha_n(\tilde{s}-\tilde{p}_{n}^*)^2
+\omega_{n}s\tilde{p}_{n}^*+\phi_{n}\tilde{s}^2+\delta_{n}|\tilde{p}_1^*=p_1^*,\cdots,\tilde{p}_{n-1}^*=p_{n-1}^*,\\&
\tilde{p}_1=p_1,\cdots,\tilde{p}_{n-1}=p_{n-1},\tilde{s}=s]\\
=&\max_x\{x[Ks-(1+\gamma_n)p_{n-1}^*-\lambda_nx]+\alpha_n[s-p_{n-1}^*-\gamma_n'p_{n-1}^*-\eta_nx]^2+\\&
\omega_ns(p_{n-1}^*+\gamma_n'p_{n-1}^*+\eta_nx)+\phi_ns^2+\delta_n\},
\end{aligned}
\end{equation}
The  first order condition is
\begin{equation}\label{eqa5}\begin{aligned}&(2\lambda_n-2\alpha_n\eta_n^2)x_n-(K-2\alpha_n\eta_n+\eta_n\omega_n)(s-p_{n-1}^*)\\
&+[1+\gamma_n-2\alpha_n\eta_n(1+\gamma_n')-K+2\alpha_n\eta_n-\eta_n\omega_n]p_{n-1}^*=0.\end{aligned}
\end{equation}
In the rational sense, $\tilde{s}-\tilde{p}_{n-1}^*$ is independent
with $\tilde{p}_{n-1}^*$. If our  suppose mixed strategy hold, then
we have
\begin{equation}\label{eqa6}\begin{aligned}&2\lambda_n-2\alpha_n\eta_n^2=0,\end{aligned}
\end{equation}
\begin{equation}\label{eqa7}\begin{aligned}&K-2\alpha_n\eta_n+\eta_n\omega_n=0,\end{aligned}
\end{equation}
\begin{equation}\label{eqa8}\begin{aligned}&
1+\gamma_n-2\alpha_n\eta_n(1+\gamma_n')=0.\end{aligned}
\end{equation}
Since
$\tilde{p}_n-\tilde{p}_{n-1}^*=\gamma_n\tilde{p}_{n-1}^*+\lambda_n(\tilde{x}_n+\tilde{\mu}_n)$,
$\tilde{p}_n^*-\tilde{p}_{n-1}^*=\gamma_n'\tilde{p}_{n-1}^*+\eta_n\tilde{x}_n$
and $E(\tilde{p}_n)=E(\tilde{p}_n^*)$, we have
\begin{equation}\label{eqa9}\gamma_n=-\lambda_n\theta_n, \gamma_n'=-\eta_n\theta_n.\end{equation}
Combing the market efficient condition with Eq. $(\ref{eqa9})$, we
have
\begin{equation}\label{eq10}\begin{aligned}\tilde{p}_n-\tilde{p}_{n-1}^*=&E_1(\tilde{v}-\tilde{p}_{n-1}^*|
\tilde{p}_1^*=p_1^*,\cdots,\tilde{p}_{n-1}^*=p_{n-1}^*,
\tilde{p}_1=p_1,\cdots,\tilde{p}_{n-1}=p_{n-1},\tilde{x}_1+\tilde{\mu}_1,\cdots,\tilde{x}_n+\tilde{\mu}_n)\\=&
E_1\left[\tilde{v}-\tilde{p}_{n-1}^*|\tilde{x}_n+\tilde{\mu}_n+\frac{\gamma_n}{\lambda_n}\tilde{p}_{n-1}^*\right]\\
=&E_1\left[\tilde{v}-\tilde{p}_{n-1}^*|\frac{\gamma_n}{\lambda_n}\tilde{p}_{n-1}^*+\beta_n(1+\gamma_n)
(\tilde{s}-\tilde{p}_{n-1}^*)+\theta_n\tilde{s}+\tilde{z}_n+\tilde{\mu}_n\right]\\
=&E_1\left[\tilde{v}-\tilde{p}_{n-1}^*|(\beta_n(1+\gamma_n)+\theta_n)
(\tilde{s}-\tilde{p}_{n-1}^*)+\tilde{z}_n+\tilde{\mu}_n\right],
\end{aligned}
\end{equation}
and
\begin{equation}\label{eqa12}\begin{aligned}\tilde{p}_n^*-\tilde{p}_{n-1}^*
=&E_1(\tilde{v}-\tilde{p}_{n-1}^*|
\tilde{p}_1^*=p_1^*,\cdots,\tilde{p}_{n-1}^*=p_{n-1}^*,
\tilde{p}_1=p_1,\cdots,\tilde{p}_{n-1}=p_{n-1},\tilde{x}_1,\cdots,\tilde{x}_n))\\
=&
E_1\left[\tilde{v}-\tilde{p}_{n-1}^*|\tilde{x}_n+\frac{\gamma_n'}{\eta_n}p_{n-1}^*\right]\\
=&E_1\left[\tilde{v}-\tilde{p}_{n-1}^*|\frac{\gamma_n'}{\eta_n}p_{n-1}^*+\beta_n(1+\gamma_n)
(\tilde{s}-p_{n-1}^*)+\theta_n\tilde{s}+z_n\right]\\
=&E_1\left[\tilde{v}-\tilde{p}_{n-1}^*|(\beta_n(1+\gamma_n)+\theta_n)
(\tilde{s}-\tilde{p}_{n-1}^*)+\tilde{z}_n\right],
\end{aligned}
\end{equation}
so
\begin{equation}\label{eqa11}\lambda_n=\frac{(\beta_n(1+\gamma_n)+\theta_n)\Sigma_{n-1}}
{(\beta_n(1+\gamma_n)+\theta_n)^2\Sigma_{n-1}+\sigma_{\mu}^2+\sigma_{z_n}^2},
\end{equation}
\begin{equation}\label{eqa13}\eta_n=\frac{(\beta_n(1+\gamma_n)+\theta_n)\Sigma_{n-1}}
{(\beta_n(1+\gamma_n)+\theta_n)^2\Sigma_{n-1}+\sigma_{z_n}^2}.
\end{equation}
Combing Eqs. $(\ref{eqa6})$ and $(\ref{eqa7})$
$$\eta_n=\frac{K}{2\alpha_n-\omega_n}, \lambda_n=\frac{K^2\alpha_n}{(2\alpha_n-\omega_n)^2},$$
Eqs. $(\ref{eqa6})$, $(\ref{eqa11})$ and $(\ref{eqa13})$ yield
\begin{equation}\frac{(\beta_n(1+\gamma_n)+\theta_n)\Sigma_{n-1}}
{(\beta_n(1+\gamma_n)+\theta_n)^2\Sigma_{n-1}+\sigma_{z_n}^2+\sigma_{\mu}^2}=\frac{K\alpha_n}{2\alpha_n-\omega_n}\frac{(\beta_n(1+\gamma_n)+\theta_n)\Sigma_{n-1}}
{(\beta_n(1+\gamma_n)+\theta_n)^2\Sigma_{n-1}+\sigma_{z_n}^2},
\end{equation}
i.e.
\begin{equation}\label{eqa14}(\beta_n(1+\gamma_n)+\theta_n)^2\Sigma_{n-1}+\sigma_{z_n}^2=\frac{K\alpha_n}
{2\alpha_n-K\alpha_n-\omega_n} \sigma_{\mu}^2.
\end{equation}
Substitute the above formula into the Eq. $(\ref{eqa11})$, it is
easy to get
\begin{equation}
\beta_n(1+\gamma_n)+\theta_n=\frac{(2\alpha_n-\omega_n)\sigma_{\mu}^2\lambda_n}{(2\alpha_n-K\alpha_n-\omega_n)\Sigma_{n-1}}
,\end{equation} and
\begin{equation}\label{eqa15}\begin{aligned}
&\sigma_{z_n}^2=\frac{K\alpha_n\sigma_{\mu}^2}{2\alpha_n-K\alpha_n-\omega_n}-
\frac{(2\alpha_n-\omega_n)^2\sigma_{\mu}^4\lambda_n^2}{(2\alpha_n-K\alpha_n-\omega_n)^2\Sigma_{n-1}}.
\end{aligned}
\end{equation}

Since
$\theta_n=\frac{2\alpha_n\eta_n-1}{\lambda_n}=\frac{2\alpha_nK-2\alpha_n+\omega_n}{(2\alpha_n-\omega_n)\lambda_n}$,
we have
\begin{equation}\label{eqa15'}\beta_n(1+\gamma_n)=\frac{K^4\alpha_n^2\sigma_{\mu}^2}{(2\alpha_n-K\alpha_n-\omega_n)(2\alpha_n-\omega-n)^3\Sigma_{n-1}\lambda_n}
-\frac{2\alpha_nK-2\alpha_n+\omega_n}{(2\alpha_n-\omega_n)\lambda_n}.
\end{equation}
Using Eqs. (\ref{eqa15'}) and the projection theorem for normally
distributed random variables, we obtain
\begin{equation}\label{eqa16}\begin{aligned}
\Sigma_n
=\Sigma_{n-1}-\frac{K\lambda_n}{2\alpha_n-K\alpha_n-\omega_n}\sigma_{\mu}^2.
\end{aligned}
\end{equation}

Since the strategy of the insider is
$\tilde{x}_n=\beta_n(1+\gamma_n)(\tilde{s}-\tilde{p}_{n-1}^*)+\theta_n\tilde{s}+\tilde{z}_n$,
we have
\begin{equation}\label{eqa17}\begin{aligned}
&E_K\{\tilde{\pi}_n|\tilde{p}_1^*=p_1^*,\cdots,\tilde{p}_{n-1}^*=p_{n-1}^*,
\tilde{p}_1=p_1,\cdots,\tilde{p}_{n-1}=p_{n-1},\tilde{s}=s\}\\
=&E_K\{[\beta_n(1+\gamma_n)(\tilde{s}-\tilde{p}_{n-1}^*)+\theta_n\tilde{s}+\tilde{z}_n][\tilde{v}-\tilde{p}_{n-1}^*
-\gamma_n\tilde{p}_{n-1}^*-\beta_n(1+\gamma_n)
\lambda_n(\tilde{s}-\tilde{p}_{n-1}^*)-\lambda_n\theta_n\tilde{s}-\lambda_n\tilde{z}_n-\lambda_n\tilde{\mu}_n]\\&
+\alpha_n[\tilde{s}-\tilde{p}_{n-1}^*-\gamma'_n
\tilde{p}_{n-1}^*-\beta_n(1+\gamma_n)
\eta_n(\tilde{s}-\tilde{p}_{n-1}^*)-\eta_n\theta_n\tilde{s}-\eta_n\tilde{z}_n]^2\\&
+\omega_{n}\tilde{s}[\tilde{p}_{n-1}^*+\gamma_n'\tilde{p}_{n-1}^*+\beta_n(1+\gamma_n)
\eta_n(\tilde{s}-\tilde{p}_{n-1}^*)+\eta_n\theta_n\tilde{s}+\eta_n\tilde{z}_n]+\phi_{n}\tilde{s}^2+\delta_{n}\\&
|\tilde{p}_1^*=p_1^*,\cdots,\tilde{p}_{n-1}^*=p_{n-1}^*,
\tilde{p}_1=p_1,\cdots,\tilde{p}_{n-1}=p_{n-1},\tilde{s}=s\}\\
=&
[(1+\gamma_n-\lambda_n\beta_n(1+\gamma_n))\beta_n(1+\gamma_n)+\alpha_n(1+\gamma_n'-\eta_n\beta_n(1+\gamma_n))^2](s-p_{n-1}^*)^2+
\\&[\beta_n(1+\gamma_n)(1-K)-(1+\gamma_n-\lambda_n\beta_n(1+\gamma_n))\theta_n
+\omega_n[1-(\eta_n\beta_n(1+\gamma_n)-\gamma_n')]]sp_{n-1}^*\\&
+[\phi_n+\beta_n(1+\gamma_n)(K-1)+(K+\gamma_n-\lambda_n\beta_n(1+\gamma_n))
\theta_n+\omega_n(\eta_n\beta_n(1+\gamma_n)-\gamma_n')]s^2+\\&
\alpha_n\eta_n^2\sigma_{z_n}^2+\delta_n-\lambda_n\sigma_{z_n}^2,
\end{aligned}
\end{equation}
thus we obtain that  $\alpha_{n-1}$, $\omega_{n-1}$, $\phi_{n-1}$
and $\delta_{n-1}$ are given by Eqs. $(\ref{eq4.5})-(\ref{eq4.8})$,
respectively.

We have proved that the difference equation system given by Eqs.
$(\ref{eq4.5})$-$(\ref{eq4.14})$  describes a linear equilibrium of
the model, next we  prove that this equilibrium is the unique linear
equilibrium with the boundary condition
$\alpha_N=\omega_N=\phi_N=\delta_N=0.$

Given $\alpha_n,$ $\omega_n$ and $\Sigma_n,$ $\lambda_n$ and
$\eta_n$ are uniquely  determined by Eqs. $(\ref{eq4.9})$ and
$(\ref{eq4.10})$. Furthermore, given $\alpha_n$, $\omega_n$,
$\phi_n$, $\delta_n$, $\Sigma_n$ and $\lambda_n$, we can obtain
$\beta_n$, $\theta_n$, $\gamma_n$, $\gamma_n'$, $\Sigma_{n-1}$,
$\alpha_{n-1}$, $\omega_{n-1}$, $\phi_{n-1}$, $\delta_{n-1}$ from
Eqs. $(\ref{eq4.5})$-$(\ref{eq4.14})$. Thus we can iterate the
system backwards one step. Given the the boundary
condition$\alpha_N=\omega_N=\phi_N=\delta_N=0,$ the backward
iteration procedure yields a family of solutions to the difference
equation system parameterized by the terminal value $\Sigma_N.$ But
$\Sigma_N$ is proportional to $\Sigma_0$ in any solution and
consequently there is a unique solution given the initial value
$\Sigma_0$.

\textbf{ Proof of Proposition \ref{pro4.2}} Note that
$\tilde{x}_n$,$\tilde{y}_n$, and $\tilde{\mu}_n$ are normally
distributed with zero mean, we have
\begin{equation}\begin{aligned}\label{a22}E_1[Vol_n]=&\frac{1}{2}(E_1[|\tilde{x}_n|]+E_1[|\tilde{y}_n|]+E_1[|\tilde{\mu}_n|])
\\
=&\frac{1}{\sqrt{2\pi}}(\sqrt{var_1(\tilde{x}_n)}+\sqrt{var_1(\tilde{\mu}_n)}+\sqrt{var_1(\tilde{y}_n)})\\
=&\frac{1}{\sqrt{2\pi}}(V_n^i+V_n^l+V_n^m),
\end{aligned}.
\end{equation}
where $V_n^i=\sqrt{var_1(\tilde{x}_n)}$,
$V_n^m=\sqrt{var_1(\tilde{y}_n)}$, and
$V_n^l=\sqrt{var_1(\tilde{\mu}_n)}$.

We only need to calculate $ var_1[\tilde{y}_n]$ and $
var_1[\tilde{y}_N]$, and the rests are similar and easy. By
definition, and Eq. (\ref{eq4.1})
\begin{equation}\begin{aligned}\label{a23}\tilde{y}_n=\tilde{x}_n+\tilde{\mu}_n =\beta_n(1+\gamma_n)(\tilde{s}-p_{n-1}^*)
+\theta_n\tilde{s}+\tilde{z}_n+\tilde{\mu}_n,
\end{aligned}
\end{equation}
for $n=1,2,\cdots,N-1,$
 and
 \begin{equation}\begin{aligned}\label{a24}
 \tilde{y}_N=\tilde{x}_N+\tilde{\mu}_N=\beta_N(1+\gamma_N)(\tilde{s}-p_{N-1}^*)+\theta_N\tilde{s}+\tilde{\mu}_N.
 \end{aligned}
\end{equation}
Taking variance on both sides of Eqs. (\ref{a23}) and (\ref{a24}),
respectively, we obtain
\begin{equation}\begin{aligned}\label{a25}
 var_1[\tilde{y}_n]=[\beta_n(1+\gamma_n)+\theta_n]^2\Sigma_{n-1}+\theta_n^2(\Sigma_0+p_0^2-\Sigma_{n-1})+\sigma_{z_n}^2+\sigma_{\mu}^2,
 \end{aligned}
\end{equation}
\begin{equation}\begin{aligned}\label{a26}
 var_1[\tilde{y}_N]=(\beta_N(1+\gamma_N)+\theta_N)^2\Sigma_{N-1}+\theta_N^2(\Sigma_0+p_0^2-\Sigma_{N-1})+\sigma_{\mu}^2.
 \end{aligned}
\end{equation}

Applying Eqs. (\ref{eq4.11}) and (\ref{eq4.12}) into
Eq.$(\ref{a25})$ yields Eq. $(\ref{eq4.19})$, and applying Eq.
$(\ref{eq4.14})$ into Eq. $(\ref{a26})$ yields Eq. $(\ref{eq4.22}).$

\textbf{ Proof of Proposition \ref{pro4.3}} Since $\forall
n=1,2,\cdots N,$
\begin{equation}\label{eq4.211}\lambda_n=a_n\frac{\sqrt{\Sigma_{n-1}}}{\sigma_{\mu}},~~
~\alpha_n=b_n\frac{\sigma_{\mu}}{\sqrt{\Sigma_{n-1}}},
~~\omega_n=c_n\frac{\sigma_{\mu}}{\sqrt{\Sigma_{n-1}}}.
\end{equation}
The boundary conditions Eqs. $(\ref{eq4.13'})$ and $(\ref{eq4.14'})$
imply $b_N=0$,  $c_N=0$, $a_N=\frac{\sqrt{K(2-K)}}{2}$.  Substitute
Eq. $(\ref{eq4.211})$ into Eqs. $(\ref{eq4.14})$, $(\ref{eq4.12})$,
$(\ref{eq4.10})$, $(\ref{eq4.11})$, $(\ref{eq4.12'})$ and
$(\ref{eq4.13})$ respectively,  we can get Eqs.
$(\ref{eq4.26})$-$(\ref{eq4.31})$. And Eq. $(\ref{eq4.9})$ implies
Eq. $(\ref{eq4.23'})$. Using Eq. $(\ref{eq4.5})$ we can obtain
\begin{equation}\label{eqa27}
b_{n-1}\left(1-\frac{Ka_{n-1}}{2b_{n-1}-Kb_{n-1}-c_{n-1}}\right)^{\frac{1}{2}}=
q_n,
\end{equation}
where $q_n$ is given by $(\ref{eq4.24})$. And using $(\ref{eq4.6})$,
we can have
\begin{equation}\label{eqa28}
b_{n-1}\left(1-\frac{Ka_{n-1}}{2b_{n-1}-Kb_{n-1}-c_{n-1}}\right)^{\frac{1}{2}}=
z_n,
\end{equation}
where $z_n$ is given by $(\ref{eq4.25})$. Using Eqs. $(\ref{eqa27})$
and $(\ref{eqa28})$, we can get $(\ref{eq4.22})$ and
$(\ref{eq4.23})$.

\makeatletter
\renewcommand{\@biblabel}[1]{}
\renewcommand{\@cite}[1]{ }
\makeatother


\begin{thebibliography}{99}
\bibitem{Admati}
 Admati, A.R., Pfleiderer, P., 1988. A theory of intraday patterns: volume and price variability.
 Review of Financial Studies 1,
3-40.

\bibitem{Back}

 Back, K., 1992.  Insider trading in continuous time. Review of Financial
 Studies 5(3), 387-409.

\bibitem{Benos}
Benos, A.V., 1998.  Aggressiveness and survival of overconfident
traders. Journal of Financial Markets 1, 353-383.

\bibitem{Fishman and Hagerty}
 Fishman, M.J., Hagerty, K.M., 1992. Insider trading and efficiency of stock prices.
  RAND Journal of Economics 23,
106-122.

\bibitem{Fos}
Foster, F. D., Viswanathan, S., 1996. Strategic trading when agents
forecast the forecasts of others. Journal of  Finance 51, 1437-1478.


\bibitem{Gong and Liu}
Gong, F., Liu, H., 2011. Inside trading, public disclosure and
imperfect competition. arXiv:1103.0894v1 [q-fin.TR].

\bibitem{Gong and zhou}
Gong, F., Zhou, D., 2010. Insider trading in the market with
rational expected price. arXiv:1012.2160v1 [q-fin.TR].


\bibitem{Harris and Raviv}
 Harris, M., Raviv, A., 1993. Differences of opinion make a
horse race. Review of Financial Studies 6, 473-506.

\bibitem{Holden and Subrahmanyam1}
Holden, C.W., Subrahmanyam, A., 1992. Long-lived private information
and imperfect competition. Journal of Finance 47, 247-270.

\bibitem{Holden and Subrahmanyam2}
Holden, C.W., Subrahmanyam, A., 1994. Risk aversion, imperfect
competition and long lived information. Economics Letters 44,
181-190.
\bibitem{Huddart, Hughes and Levine}
Huddart, S., Hughes, J.S. and Levine C.B., 2001. Public disclosure
and dissimulation of insider traders. Econometrica, Vol.69,  No.3,
665-681.


\bibitem{Jain and Mirman}
 Jain, N., Mirman, L.J., 1999. Insider trading with correlated signals. Economic Letters 65, 105-113.


\bibitem{Kyle}
Kyle, A.S., 1985. Continuous auctions and insider trading.
Econometrica 53, 1315-1335.

\bibitem{Kyle and wang}
Kyle, A.S., Wang, F.A., 1997. Speculation duopoly with agreement to
disagree: can overconffidence survive the market test? Journal of
Finance 52, 2073-2090.

\bibitem{odean}
Odean, T., 1998. Volume, volatility, price and profit when all
traders are above average. Journal of Finance 53, 1887-1934.
\bibitem{Shunlong Luo} Luo, S., 2001. The impact of public information on insider trading.
Economics Letters 70, 59-68.

\bibitem{Roc}
Rochet, J.C., Vila, J.L., 1994.
 Insider trading without normality. Review of Economic Studies 61, 131-152.




\bibitem{wang F.A.}

 Wang F.A., 1998. Strategic trading, asymmetric information and
 heterogeneous prior beliefs. Journal of Financial Markets 1,
 321-352.
\bibitem{wang F.A.b}
Wang, F.A., 1997. Overconffidence, delegated fund management, and
survival. Working paper, Columbia University, New York.

\bibitem{zhang a}
Zhang, W.D., 2004. Risk aversion, public disclosure and long-lived
information. Economic Letters, 85, 327-334.

\bibitem{zhang b}
Zhang, W.D., 2008. Impact of outsiders and disclosed insider trades,
Economic Letters, 5, 137-145.

\end{thebibliography}
\end{document}